\documentclass[onecolumn,aps,pre]{revtex4}

\usepackage{graphicx}

\newcommand{\middlefig}{.2\textwidth}

\begin{document}

\title{Motion of discrete solitons assisted by nonlinearity management}
\author{Jes\'us Cuevas$^1$, Boris A. Malomed$^{2}$ and P.G.\ Kevrekidis$^{3}$}
\affiliation{$^{1}$ Grupo de F{\'i}sica No Lineal, Departamento de
F{\'i}sica Aplicada I, Escuela Universitaria Polit{\'{e}}cnica, C/
Virgen de {\'{A}}frica, 7, 41011 Sevilla, Spain\\
$^{2}$ Department of Interdisciplinary Studies, School of Electrical
Engineering, Faculty of Engineering, Tel Aviv University, Tel Aviv 69978,
Israel\\
$^{3}$ Department of Mathematics and Statistics, University of
Massachusetts, Amherst MA 01003-4515, USA}

\begin{abstract}
We demonstrate that periodic modulation of the nonlinearity
coefficient in the discrete nonlinear Schr\"{o}dinger (DNLS)\
equation can strongly facilitate creation of traveling solitons in
the lattice. We predict this possibility in an analytical form,
and test it in direct simulations. Systematic simulations reveal
several generic dynamical regimes, depending on the amplitude and
frequency of the time modulation, and on initial thrust which sets
the soliton in motion. These regimes include irregular motion,
regular motion of a decaying soliton, and regular motion of a
stable one. The motion may occur in both the straight and reverse
directions, relative to the initial thrust. In the case of stable
motion, extremely long simulations in a lattice with periodic
boundary conditions demonstrate that the soliton keeps moving
as long as we can monitor
without any visible loss. Velocities of moving
stable solitons are in good agreement with the analytical
prediction, which is based on requiring a resonance between the ac
drive and motion of the soliton through the periodic potential.
%Actually, this is the first example of the ac-driven progressive
%resonant motion of a \emph{nontopological} soliton.
All the
generic dynamical regimes are mapped in the model's parameter
space. Collisions between moving stable solitons are briefly
investigated too, with a conclusion that two different outcomes
are possible: elastic bounce, or bounce with mass transfer from
one soliton to the other. The model can be realized experimentally
in a Bose-Einstein condensate trapped in a deep optical lattice.
\end{abstract}

\date{\today}
\maketitle

\bigskip

%\textit{Introduction}

\section{Introduction}

The discrete nonlinear Schr\"{o}dinger
(DNLS)\ equation is a well-known fundamental model of nonlinear
lattice dynamics, which allows to study many features of the
nonintegrable dynamics in the universal setting \cite{Panos}.
Simultaneously, this model finds direct applications to arrays of
nonlinear optical waveguides (as it was predicted long ago
\cite{ChrJos} and demonstrated in detail more recently, see Refs.
\cite{Silberberg} and references therein), and to arrays of
droplets in Bose-Einstein condensates (BECs) trapped in a very
deep optical lattice \cite{BEC}.

In all these contexts, discrete solitons are the most fundamental
dynamical excitations supported by the DNLS equation. The dynamics
of standing solitons, pinned by the underlying lattice, is
understood quite well, in terms of both numerical simulations and
analytical approximations, the most general one being based on the
variational method \cite{Michael}. However,\ moving discrete
solitons is a more complex issue \cite{moving,Amazon}. While,
strictly speaking, exact solutions for moving solitons should not
exist because of the radiation loss,
direct simulations indicate that a soliton may move freely if its
``mass" ($l^2$ norm) does not exceed a certain critical value
\cite{Amazon}. In the quasi-continuum approximation, the source of
the braking force acting on the moving soliton is the effective
Peierls-Nabarro (PN) potential induced by the lattice \cite{KM}.

In the case of the DNLS equation describing an array of nearly isolated
droplets of a BEC in a deep optical lattice, an interesting possibility is
to apply the \textit{ac Feshbach-resonance management} (FRM) to it, as it
was recently proposed in Ref. \cite{Tsoy}. FRM may be induced
by an external ac magnetic field, which periodically (in time) changes the
sign of the nonlinearity by dint of the FR affecting collisions between
atoms (for a one-dimensional BEC without the optical lattice, the concept of
FRM was proposed in Ref. \cite{FRM}).

In this work, our aim is to demonstrate that the FRM, applied
to the DNLS model, can strongly facilitate the
motion of discrete solitons (Ref.
\cite{Tsoy} was only dealing with standing solitons). The model is based on
the equation

\begin{equation}
i\dot{u}_{n}+u_{n+1}+u_{n-1}-2u_{n}+g(t)|u_{n}|^{2}u_{n}=0,  \label{dyn}
\end{equation}where $u_{n}(t)$ are the mean-field BEC\ wave functions at the lattice
sites, and the real time-dependent nonlinear coefficient, proportional to
the scattering length, is
\begin{equation}
g(t)=g_{\mathrm{dc}}+g_{\mathrm{ac}}\sin (\omega t).  \label{g}
\end{equation}
Our presentation will be structured as follows:
In section II, approximating the soliton by a
\textit{Gaussian},\textit{\ }we present an analytical estimate of
the effective PN potential for the \textit{moving} discrete
soliton. The estimate suggests that the ac modulation of $g(t)$
may indeed help to suppress the PN potential, and thus facilitate
free motion of discrete solitons. In section III, we display results of
systematic simulations, summarized in the form of diagrams in the
parameter plane $\left( \omega ,g_{\mathrm{ac}}\right) $. The
diagrams feature several generic dynamical regimes, including a
large area of stable progressive motion, that can last
indefinitely long. Collisions between solitons moving in opposite
directions are briefly considered too, with a conclusion that they
bounce from each other, sometimes featuring mass transfer between
the solitons. Diverse dynamical regimes predicted in this work
suggest straightforward possibilities for new experiments in the
BEC trapped in a deep optical lattice. Finally, in section IV,
we summarize our findings and present our conclusions.

%\textit{Analytical approximation}

\section{Analytical Approximation}

The continuum limit suggests the
following \textit{ansatz} for a moving soliton \cite{Anderson},
\begin{equation}
u_{\mathrm{ans}}(n,t)=A\exp \left[ -a\left( n-\xi (t)\right)
^{2}+i\phi (t)+(i/2)\dot{\xi}n-\left( i/4\right) \int \left(
\dot{\xi}(t)\right) ^{2}dt\right] ,  \label{Gauss}
\end{equation}where $A,a,\xi (t),$ and $\phi $ are, respectively, the amplitude, squared
inverse width, central coordinate, and phase of the soliton.
Accordingly, $\dot{\xi}$ is the soliton's velocity, $\dot{\xi}/2$
simultaneously being the wavenumber of the wave field carrying the
moving soliton. For the true soliton, in the
continuum limit, the variational approximation yields the
following relations:\begin{equation}
\dot{\phi}=3a,~A^{2}=4\sqrt{2}a/g,  \label{soliton}
\end{equation}if $g=$\textrm{$c$}$\mathrm{onst}>0$. In Eqs. (\ref{soliton}), $a$ is
regarded as an arbitrary positive constant (intrinsic parameter of the
soliton family).

To estimate the PN potential acting on the \emph{quasi-continuum} soliton in
the \emph{discrete} system, we use the Hamiltonian corresponding to the DNLS
equation (\ref{dyn}),
\begin{equation}
H=\sum_{n=-\infty }^{+\infty }\left[ 2|u_{n}|^{2}-\left( u_{n}^{\ast
}u_{n+1}+u_{n}u_{n+1}^{\ast }\right) -\frac{g}{2}\left\vert u_{n}\right\vert
^{4}\right] ,  \label{H}
\end{equation}the asterisk standing for the complex conjugation.
Substituting the ansatz (\ref{Gauss}) into $H$, the potential of
the soliton-lattice interaction can be calculated in the form of a
Fourier series, $H(\xi ,\dot{\xi})=\sum_{m=0}^{\infty
}H_{m}(\dot{\xi})\cos \left( 2\pi m\xi \right) $. In the case of a
broad soliton, for which the ansatz (\ref{Gauss}) is relevant, we
keep only the lowest harmonic ($m=1$), which is actually the PN
potential $U_{\mathrm{PN}}$.
% per se.
After some simple algebra, we
thus find
\begin{eqnarray}
U_{\mathrm{PN}}(\xi ,\dot{\xi}) &=&\frac{1}{2}\sqrt{\frac{\pi }{a}}A^{2}\exp
\left( -\frac{\pi ^{2}}{4a}\right) \left\{ 4\sqrt{2}\exp \left( -\frac{\pi
^{2}}{4a}\right) \right.  \nonumber \\
&&\left. \left[ 1+e^{-a/2}\cos \left( \dot{\xi}/2\right) \right]
-gA^{2}\sqrt{\frac{\pi }{a}}\right\} \cos \left( 2\pi \xi \right)
.  \label{PN}
\end{eqnarray}A novel feature in this estimate, in comparison with known perturbative
results \cite{KM}, is the dependence on the soliton's velocity, $\dot{\xi}$.

If the soliton's relation between $a$ and $A^{2}$, as given by Eq.
(\ref{soliton}) (for constant $g$), is substituted in Eq.
(\ref{PN}), the coefficient in front of $\cos \left( 2\pi \xi
\right) $, i.e., the amplitude of the PN potential, never
vanishes. However, it may vanish if the underlying pulse
(\ref{Gauss}) is considered not as a soliton, but just as a pulse
with $A^{2}$ and $a$ taken independently; then, the condition of
the vanishing of the PN potential determines a \emph{discrete
spectrum} of the velocities $\dot{\xi}$, in the form
\begin{equation}
1+\exp \left( -a/2\right) \cos \left( \dot{\xi}/2\right) =\left(
gA^{2}/4\right) \sqrt{\pi /\left( 2a\right) }\exp \left( \pi
^{2}/(4a)\right) ,  \label{vanishing}
\end{equation}provided that $gA^{2}$ is small enough to make the right-hand
side of Eq. (\ref{vanishing}) smaller than $2$ [this caveat is essential,
as the factor $\exp \left( \pi ^{2}/(4a)\right) $ may be exponentially large].

In this work, however, our objective is not to verify this
possibility, but rather to consider the case when the nonlinear
coefficient $g$ is a function of time, as per Eq. (\ref{g}). Note
that, for a broad soliton (small $a$), the PN potential barrier is
exponentially small, hence the soliton's kinetic energy may be
much larger than the potential. This implies that the velocity of
the soliton moving through the potential (\ref{PN}) with the
period $L=1$ contains a constant (dc) part and a small ac
correction to it, with the frequency $2\pi \dot{\xi}_{0}/L\equiv
2\pi \dot{\xi}_{0}$ \cite{peyrard}:
\begin{equation}
\dot{\xi}(t)\approx \dot{\xi}_{0}+\dot{\xi}_{1}\cos \left( 2\pi
\dot{\xi}_{0}t\right) ,\dot{\xi}_{1}^{2}\ll \dot{\xi}_{0}^{2}.
\label{velocity}
\end{equation}Then, the substitution of the expression (\ref{velocity}) into the condition
(\ref{vanishing}), which provides for the suppression of the PN
potential, one can expand its left-hand side,
\begin{equation}
1+\exp \left( -a/2\right) \cos \left( \dot{\xi}/2\right) \approx
1+\exp \left( -a/2\right) \left[ \cos \left(
\dot{\xi}_{0}/2\right) -(\dot{\xi}_{1}/2)\sin \left(
\dot{\xi}_{0}/2\right) \cos \left( 2\pi \dot{\xi}_{0}t\right)
\right] .  \label{expansion}
\end{equation}Now, substituting the variable $g(t)$ from Eq. (\ref{g}) into the right-hand
side of Eq. (\ref{vanishing}), it is obvious that
$g_{\mathrm{dc}}$ and $g_{\mathrm{ac}}$ can be chosen so as to
provide for the fulfilment of the condition, provided that the
average soliton's velocity takes the \textit{resonant value},
$\dot{\xi}_{0}=\omega /2\pi $. More generally, due to anharmonic
effects, one may expect the existence of a spectrum of resonant
velocities,\begin{equation} \dot{\xi}_{0}=\left(
c_{\mathrm{res}}\right) _{N}^{(M)}\equiv M\omega /2\pi N,
\label{res}
\end{equation}with integers $M$ and $N$.

Actually, an ac drive can support stable progressive motion of solitons at
the resonant velocities (\ref{res}) (assuming the spatial period $L=1$),
even in the presence of dissipation, in a broad class of systems. This
effect was first predicted for discrete systems (of the Toda-lattice and
Frenkel-Kontorova types) in Refs. \cite{Bonilla}, and demonstrated
experimentally in an LC electric transmission line in Ref. \cite{Kuusela}.
Later, the same effect was predicted \cite{Bob} and demonstrated
experimentally \cite{Lyosha} in continuous long Josephson junctions with a
spatially periodic inhomogeneity.
%However, a qualitative difference of the
%present prediction is that it applies to \emph{nontopological solitons},
%while all the previously known examples involved \textit{kinks}, i.e.,
%discrete of continuum solitons whose topological charge directly coupled to
%the driving field.
All of the above examples were in the context topological (kink-like)
excitations in the aforementioned models. The only example similar
to what is suggested here in the context of {\it non-topological
solitons} that we are aware of,
was in the work of \cite{nistaz} (the context in the latter
case involved damping and external, fixed and localized in space ac-drive).
Note also that the mechanism of the ac-driven motion
considered here is different from that in ratchet systems (see, e.g., Ref.
\cite{Ustinov} and references therein).

%\textit{Identification of basic dynamical regimes}

\section{Numerical Results}

We now proceed to examine the above analytical predictions through
direct numerical simulations.
We integrate Eq. (\ref{dyn}) with an initial
configuration in the form of a standing-soliton solution for the
case of $g_{\mathrm{ac}}=0$. This solution has the form of
$u_{n}(t)=v_{n}\exp (i\nu t)$, with the real field $v_{n}$ obeying
the equation
\begin{equation}
\nu v_{n}=v_{n+1}+v_{n-1}-2v_{n}+g_{\mathrm{dc}}v_{n}^{3}.  \label{nu}
\end{equation}Equation (\ref{nu}) was solved by means of well-known methods
(starting from
the anti-continuum limit). Then, to set the soliton in motion, it
was provided with a lattice momentum $q$, so that the initial condition
was\begin{equation} u_{n}(0)=v_{n}\exp (inq/2).  \label{initial}
\end{equation}Equation (\ref{initial}) implies that the soliton
will move to the right if $q>0$.

The results will be displayed for $g_{\mathrm{dc}}=1$, $\nu =1$,
and three values of the initial thrust, $q=0.25$, $q=0.5$ and
$q=1$, as these cases were found to represent a generic situation
in the plane of the ac-drive's parameters $\left( \omega
,g_{\mathrm{ac}}\right) $. Simulations were run in the interval
$0<t<100\times (2\pi /\omega )$ or longer, by means of the
fourth-order Runge-Kutta algorithm, with the time step $\Delta
t=0.002$. In\ most cases, the lattice with 251 sites was used.
Edge absorbers were installed by adding the loss term $i\gamma
u_{n}$, with $\gamma =1$, to Eq. (\ref{dyn}) at the ten sites adjacent
to each edge. Besides that, extremely long simulations were
performed in a longer lattice with periodic boundary conditions,
to verify if the motion could last for very long times, and also to
examine collisions between the solitons, see below.

If $g_{\mathrm{ac}}=0$, the soliton pushed as per Eq.
(\ref{initial}) with $q\lesssim 0.7$ does not move. Instead, it
remains pinned to the lattice, with its center oscillating around
an equilibrium position. This observation may be explained by the
fact that the kinetic energy given to the soliton is smaller than
the height of the PN potential barrier.

\begin{figure}[tbp]
\begin{center}
\begin{tabular}{ccc}
(a) & (b) & (c) \\
\includegraphics[width=\middlefig]{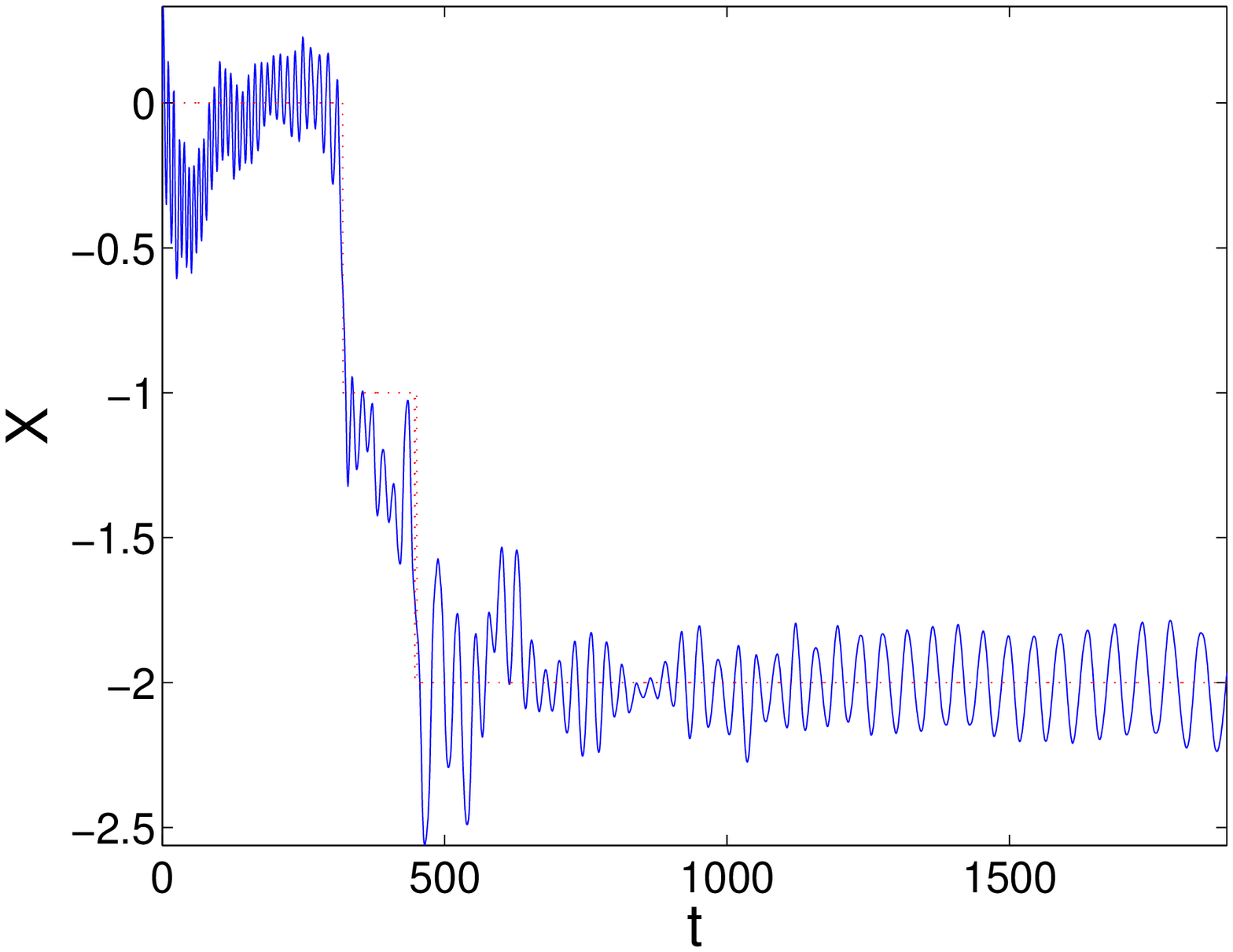} & \includegraphics[width=\middlefig]{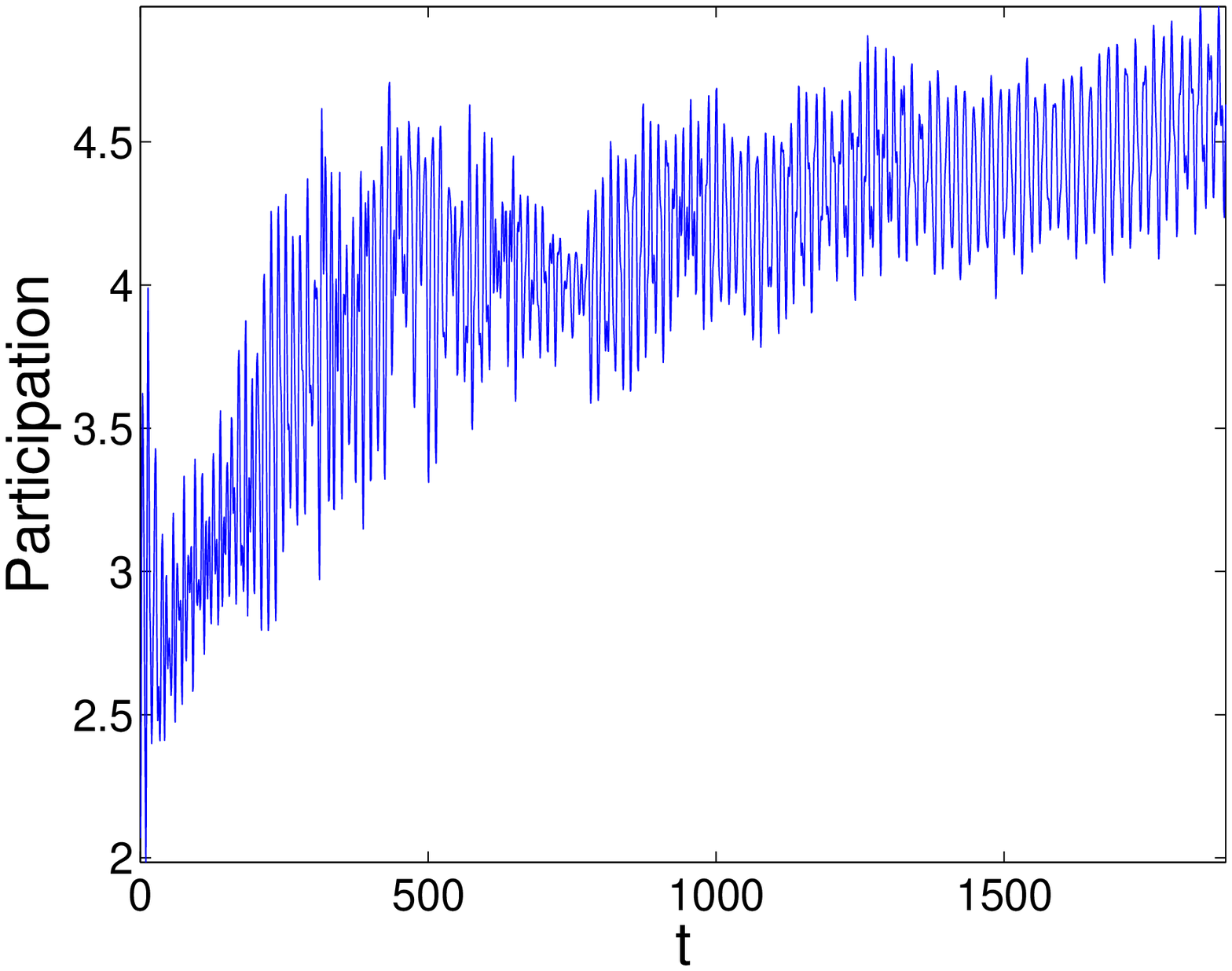} &
\includegraphics[width=\middlefig]{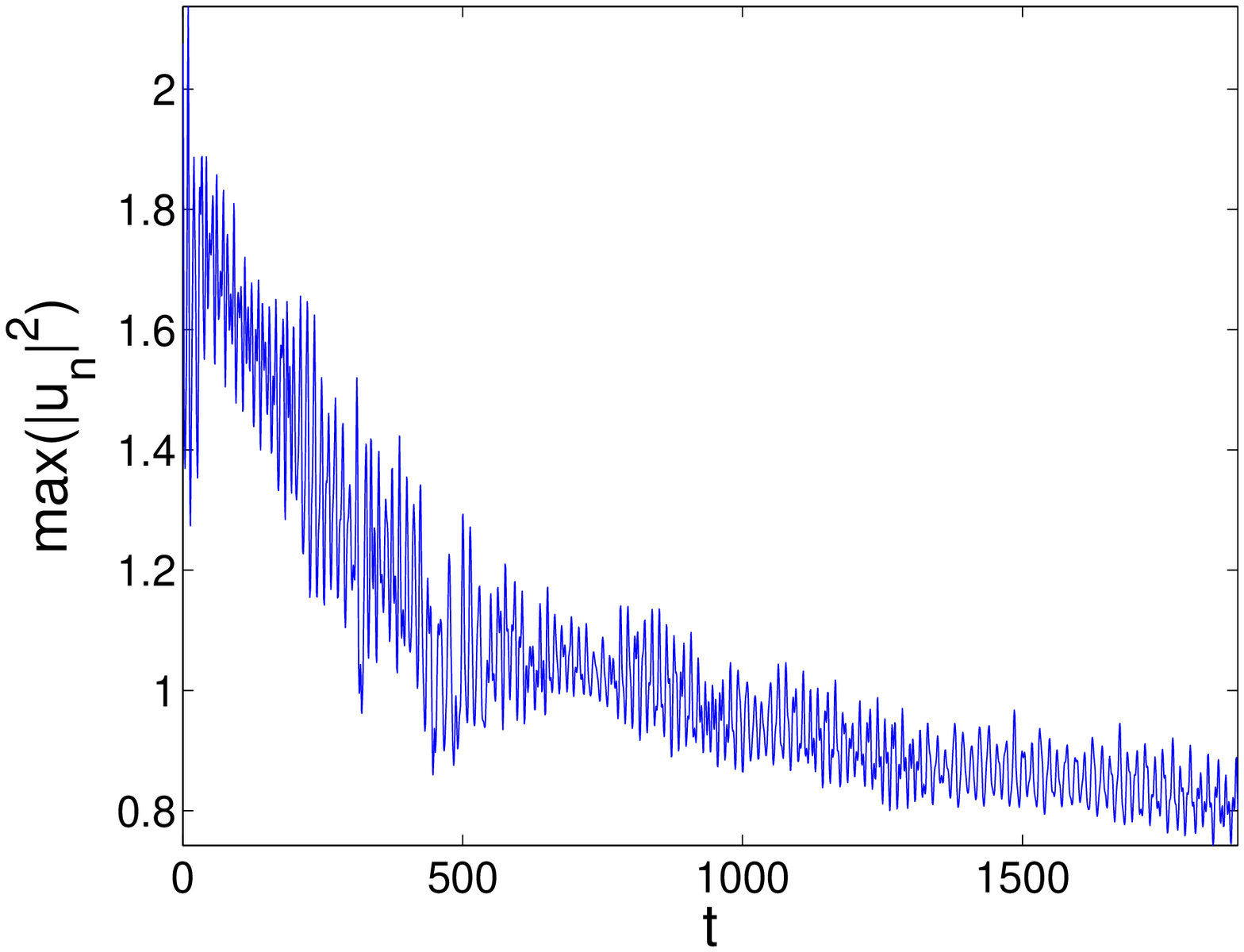}
\end{tabular}
\end{center}
\caption{A typical example of a dynamical regime in which the
soliton remains pinned, for $g_{\mathrm{ac}}=0.03$, $\protect\omega
=1$ and $\protect q =0.5$. Here and in similar plots below, panels
display the following variables as functions of time: (a) positions
of the center of mass $X$ (red) and maximum $X_{p}$ (blue) of the
density field, $|u_{n}|^{2}$; (b) the ``participation number" $P$
(which characterizes the width of the soliton); (c) the density
amplitude of the soliton, i.e., $|u_{n}|^{2}$ at $n=X_{p}$.}
\label{fig1}
\end{figure}

Several distinct types of the dynamics were observed at
$g_{\mathrm{ac}}>0$, depending on the driving frequency $\omega $
and the thrust parameter $q$. First, the soliton may remain
pinned, as shown in Fig. \ref{fig1}. In this case, the simulations
[run in the interval $0<t<300\times (2\pi /\omega )$] demonstrate
that the soliton stays pinned within a few sites from its initial position.
The central coordinate and ``participation number" $P$ (actually,
it measures an average width of the soliton), shown in Fig. \ref{fig1}
and elsewhere, are defined as\begin{equation}
X=\sum_{n}n|u_{n}|^{2}{\Big/}\sum\limits_{n}|u_{n}|^{2},~P=\left(
\sum_{n}|u_{n}|^{2}\right) ^{2}{\Large
/}\sum\limits_{n}|u_{n}|^{4}. \label{XP}
\end{equation}

\begin{figure}[tbp]
\begin{center}
\begin{tabular}{ccc}
(a) & (b) & (c) \\
\includegraphics[width=\middlefig]{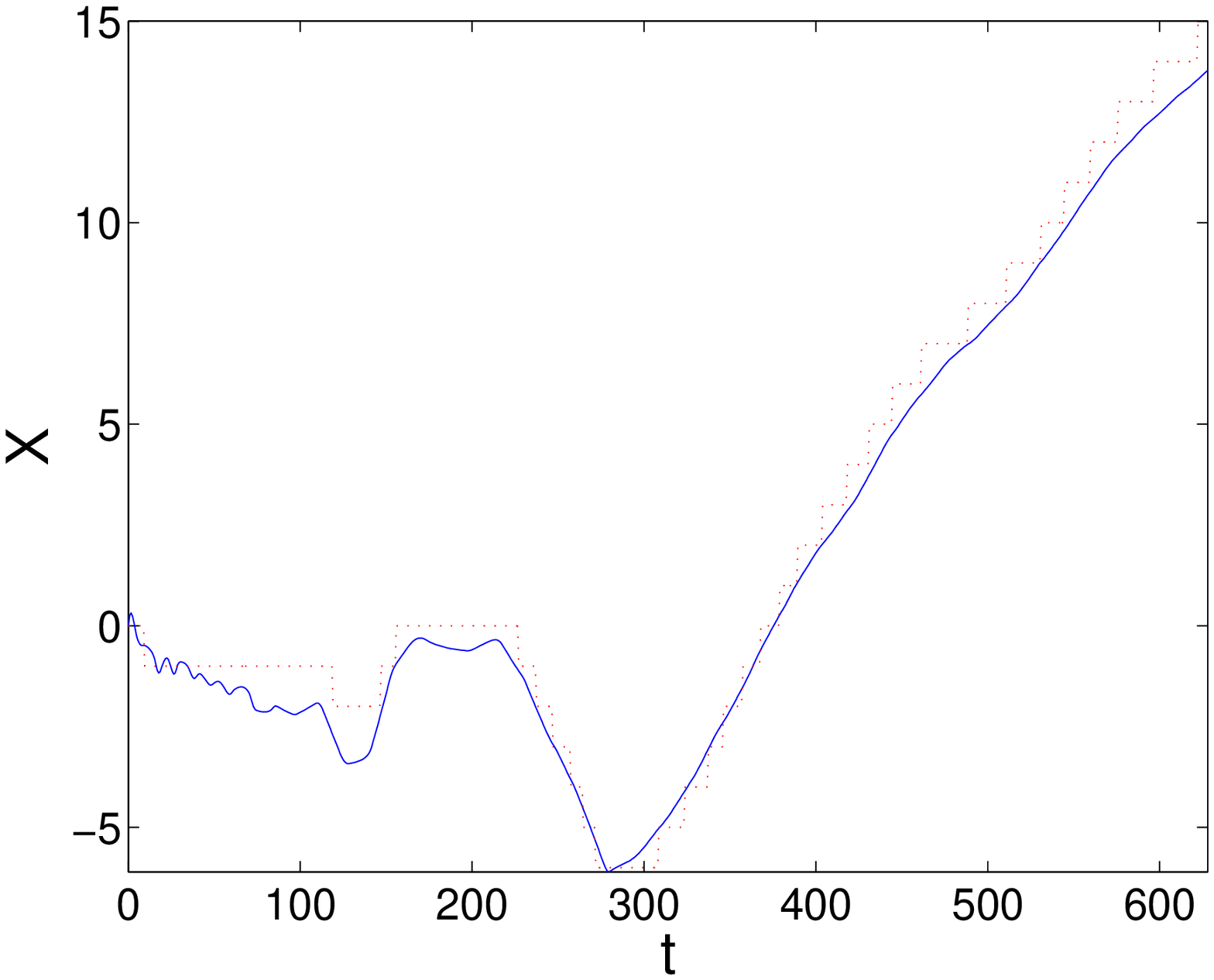} & \includegraphics[width=\middlefig]{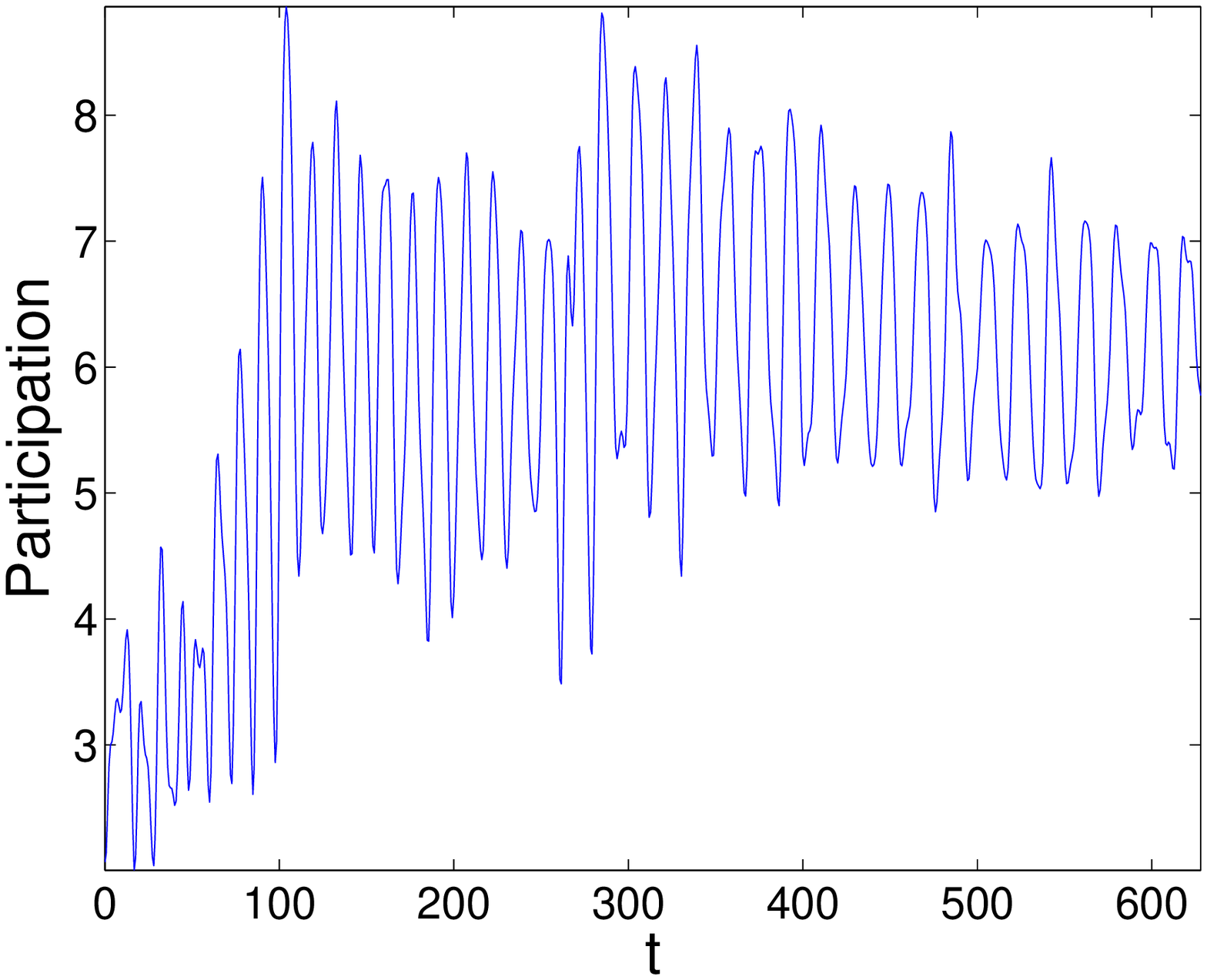} &
\includegraphics[width=\middlefig]{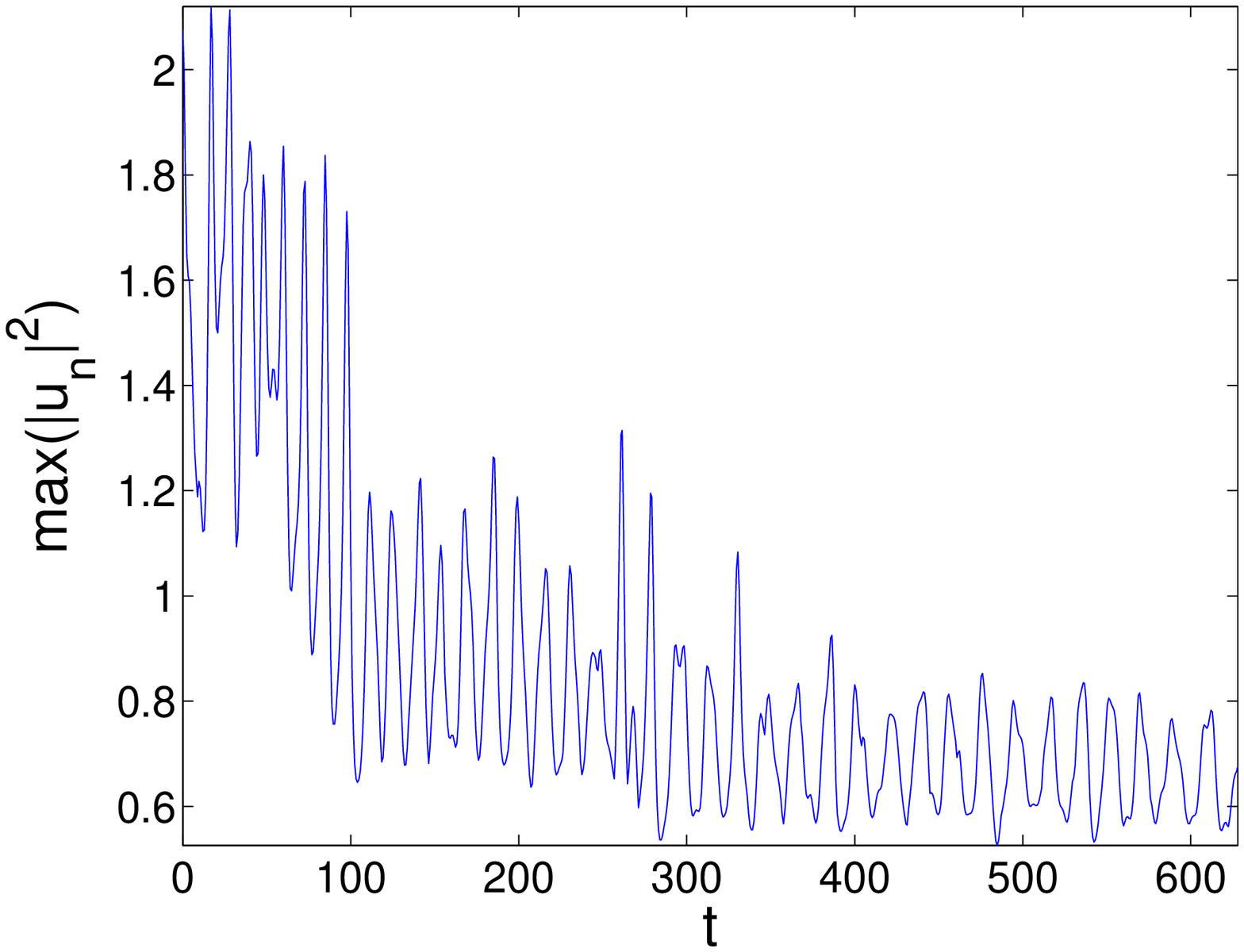}
\end{tabular}
\end{center}
\caption{A generic example of irregular motion of the soliton, for
$g_{\mathrm{ac}}=0.065$, $\protect\omega =1$ and $\protect q =0.5$.}
\label{fig2}
\end{figure}

The next generic regime is that of irregular motion, as shown in
Fig. \ref{fig2}. A characteristic feature of this regime is that
the soliton randomly changes the direction of motion several
times, and the velocity remains very small in comparison with
regimes of ``true motion", see below.

\begin{figure}[tbp]
\begin{center}
\begin{tabular}{ccc}
(a) & (b) &  \\
\includegraphics[width=\middlefig]{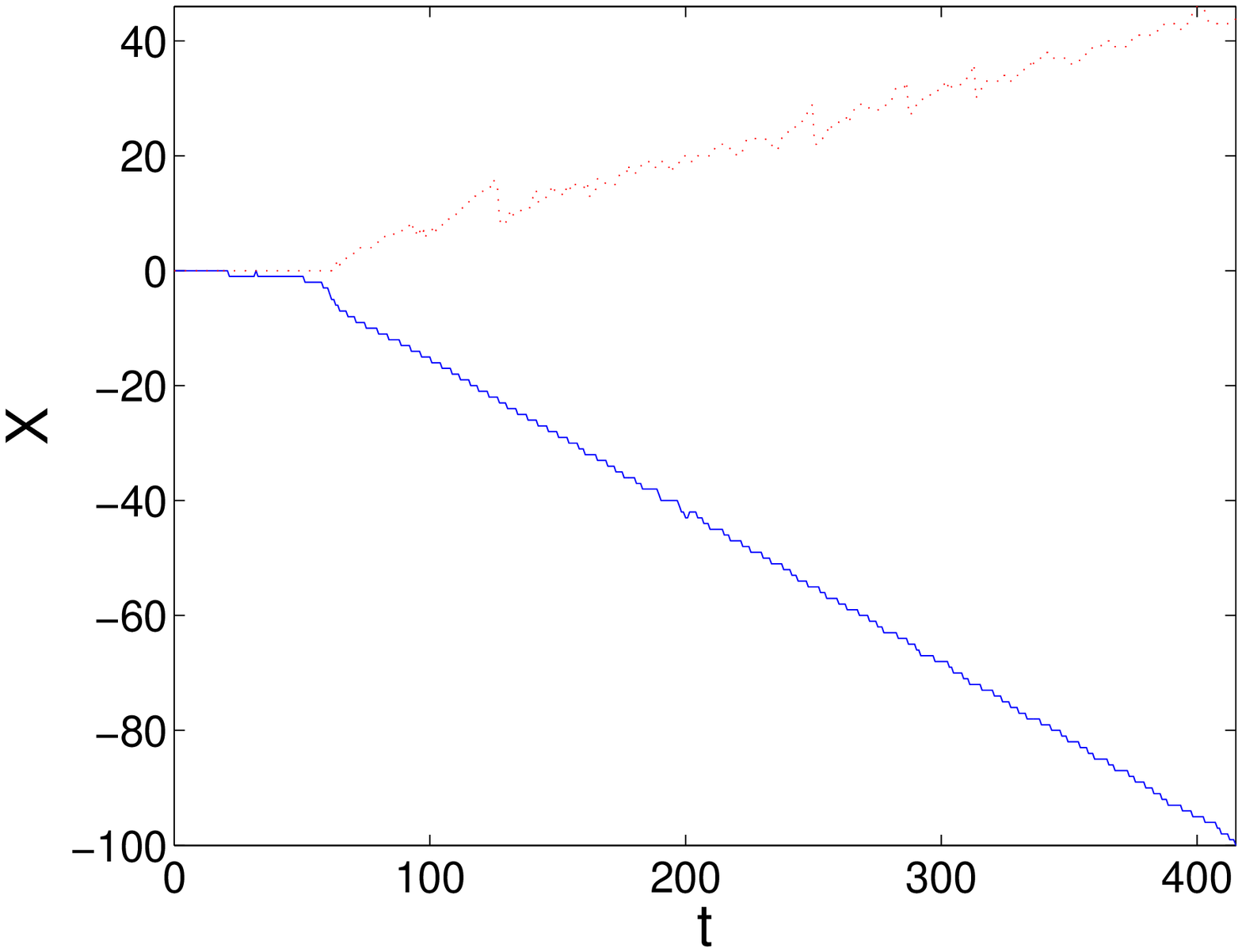} & \includegraphics[width=\middlefig]{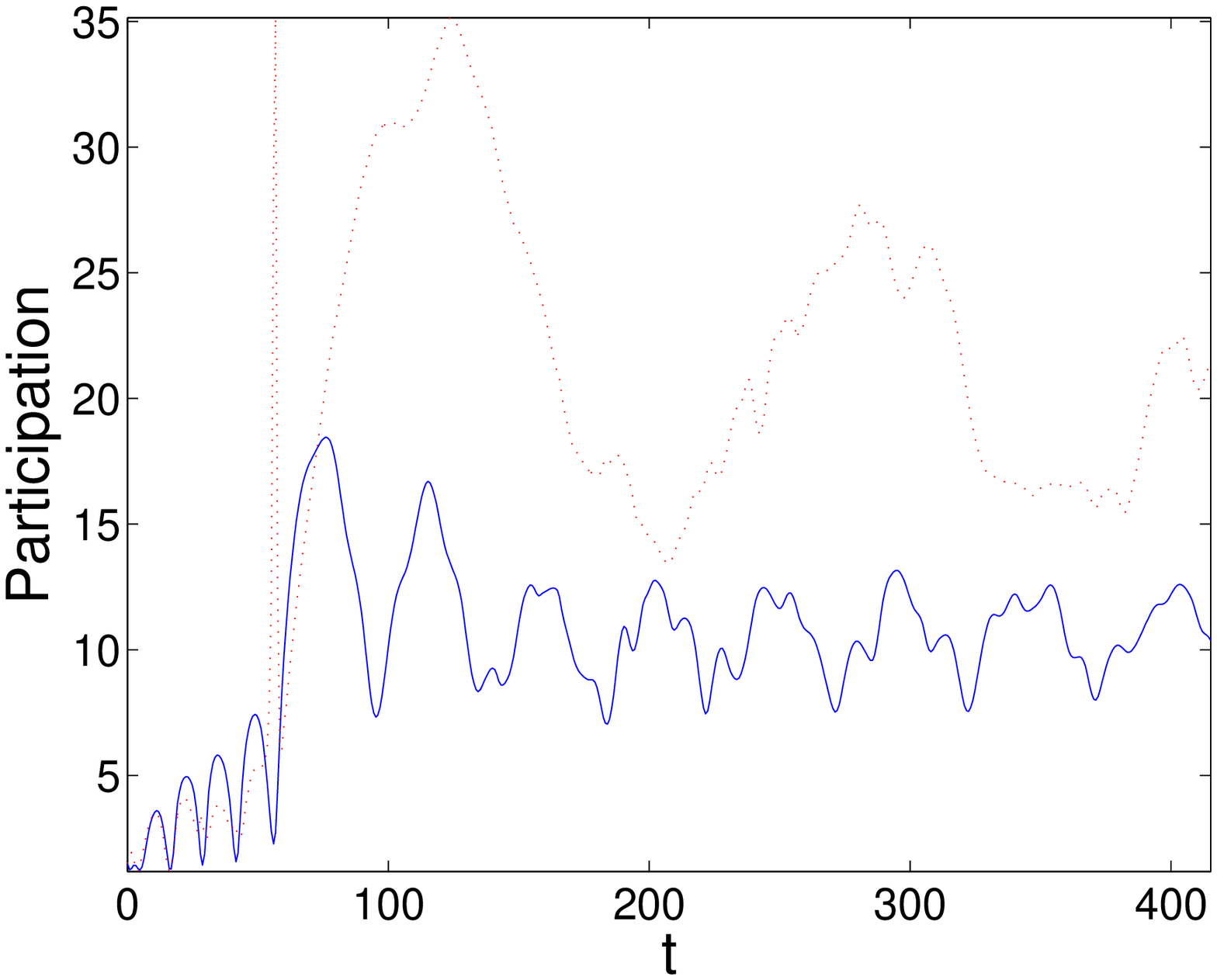} &  \\
(c) & (d) &  \\
\includegraphics[width=\middlefig]{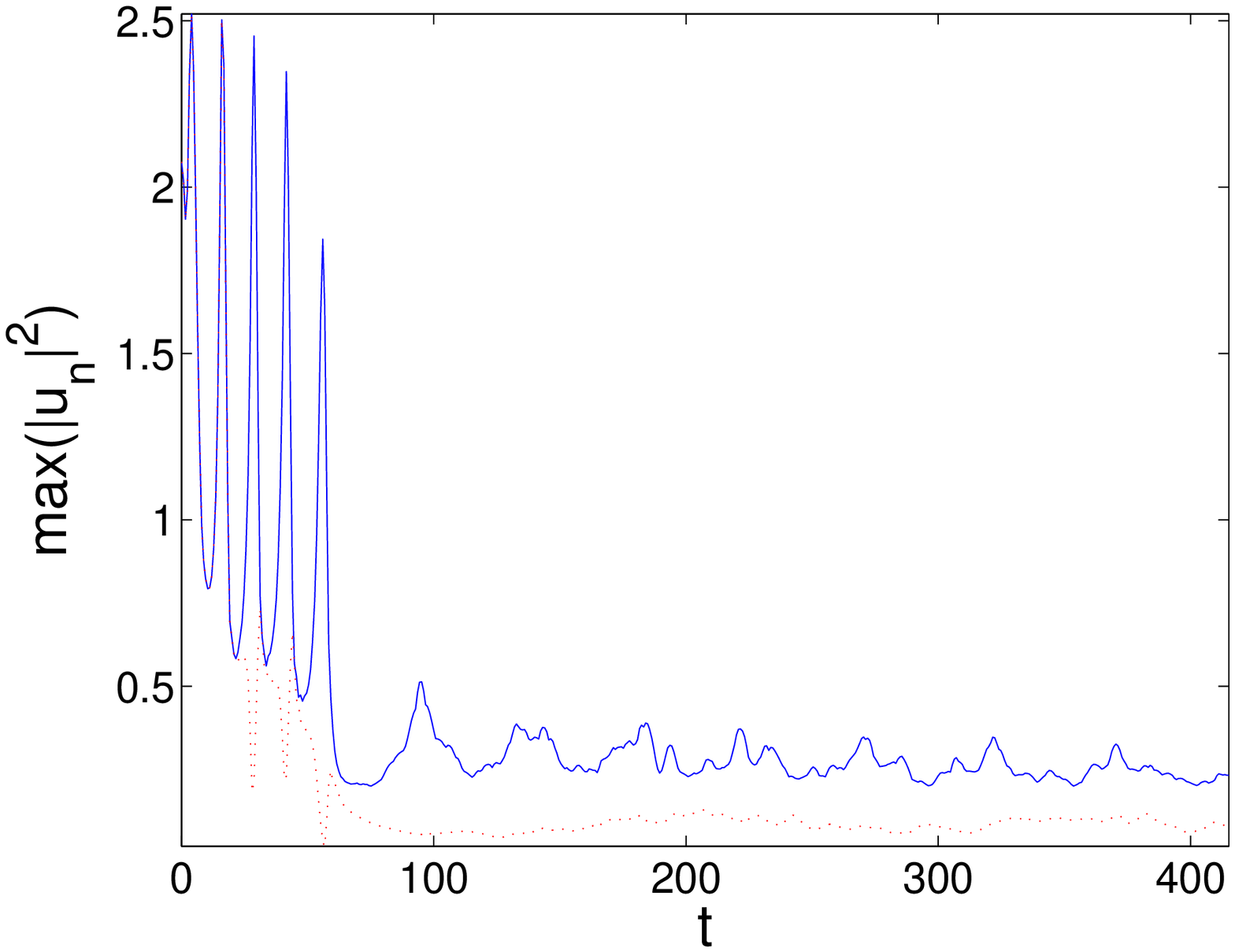} & \includegraphics[width=\middlefig]{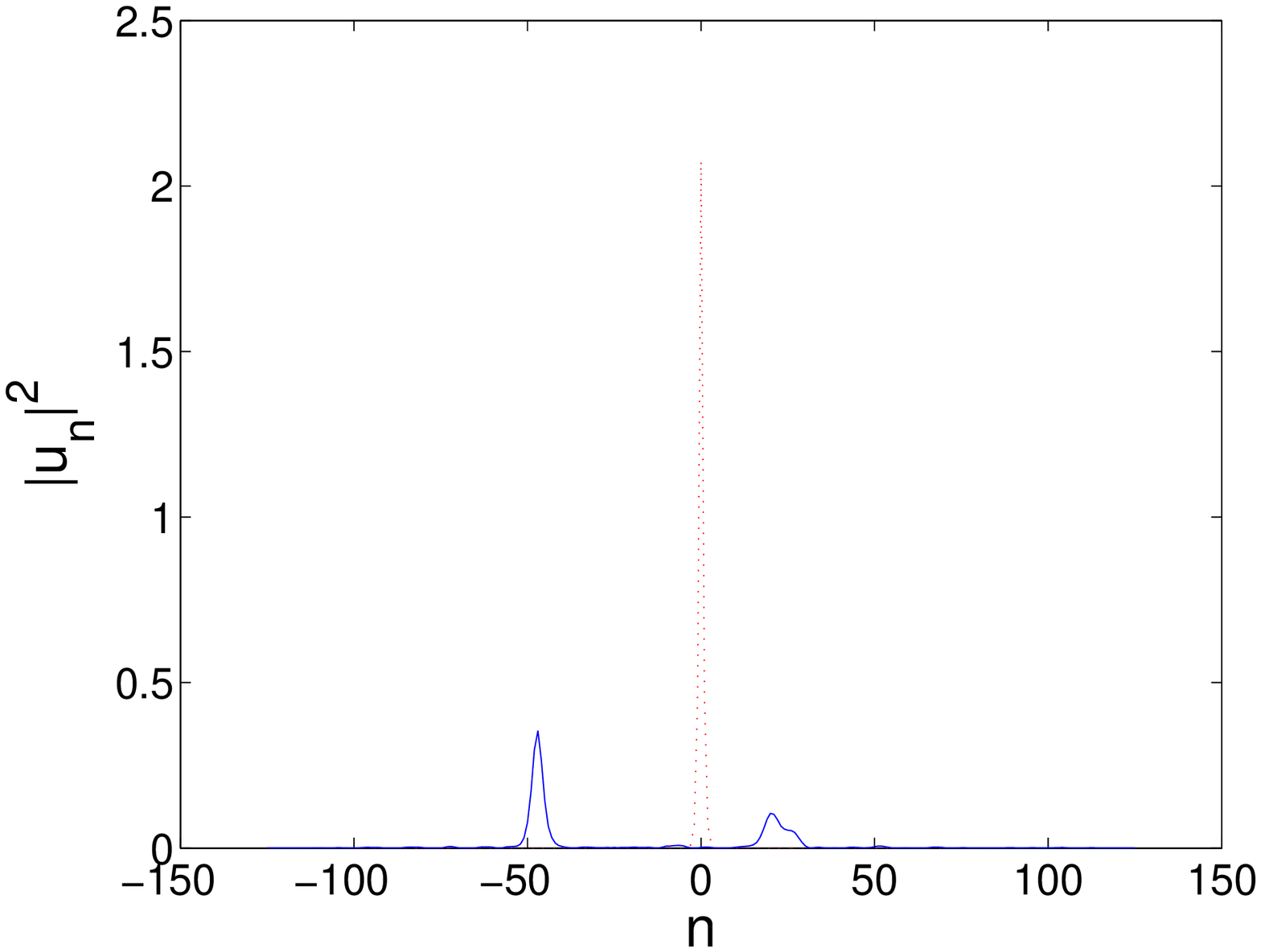} &
\end{tabular}\end{center}
\caption{An example of asymmetric splitting of the soliton, for
$g_{\mathrm{ac}}=0.196$, $\protect\omega =0.5$ and $\protect q
=0.5$. Unlike the previous figures, here panels (a) and (c) show,
respectively, the position and magnitude of two local density maxima
corresponding to the secondary solitons (splinters) past the
splitting point, and panel (b) shows, accordingly, two
``participation numbers", computed as per Eq. (\protect\ref{XP}),
but separately for $n\leq 0$ and $n\geq 0$. In this figure and
below, an additional panel (d) shows the initial and final density
distributions, $|u_{n}(t=0)|^{2}$ and
$|u_{n}(t=t_{\mathrm{fin}}|^{2}$, respectively, by the dashed and
continuous lines. In this case, $t_{\mathrm{fin}}=300$.}
\label{fig3}
\end{figure}

Under the action of a stronger drive, the soliton can also split into two
secondary ones moving in opposite directions, see Fig. \ref{fig3}. As is
seen, the splitting is strongly asymmetric, and the heavier secondary
soliton (splinter) may move both forward and backward, with respect to the
initial push. We notice that splitting of a \emph{quiescent} soliton
(without any initial thrust applied to it) into two \emph{symmetric}
splinters, moving in opposite directions with equal velocities, was
reported, in the same model based on Eq. (\ref{dyn}), in Ref. \cite{Tsoy}.
The initial push applied to the soliton is a natural cause for the symmetry
breaking observed here in the case of the splitting.

\begin{figure}[tbp]
\begin{center}
\begin{tabular}{cc}
(a) & (b) \\
\includegraphics[width=\middlefig]{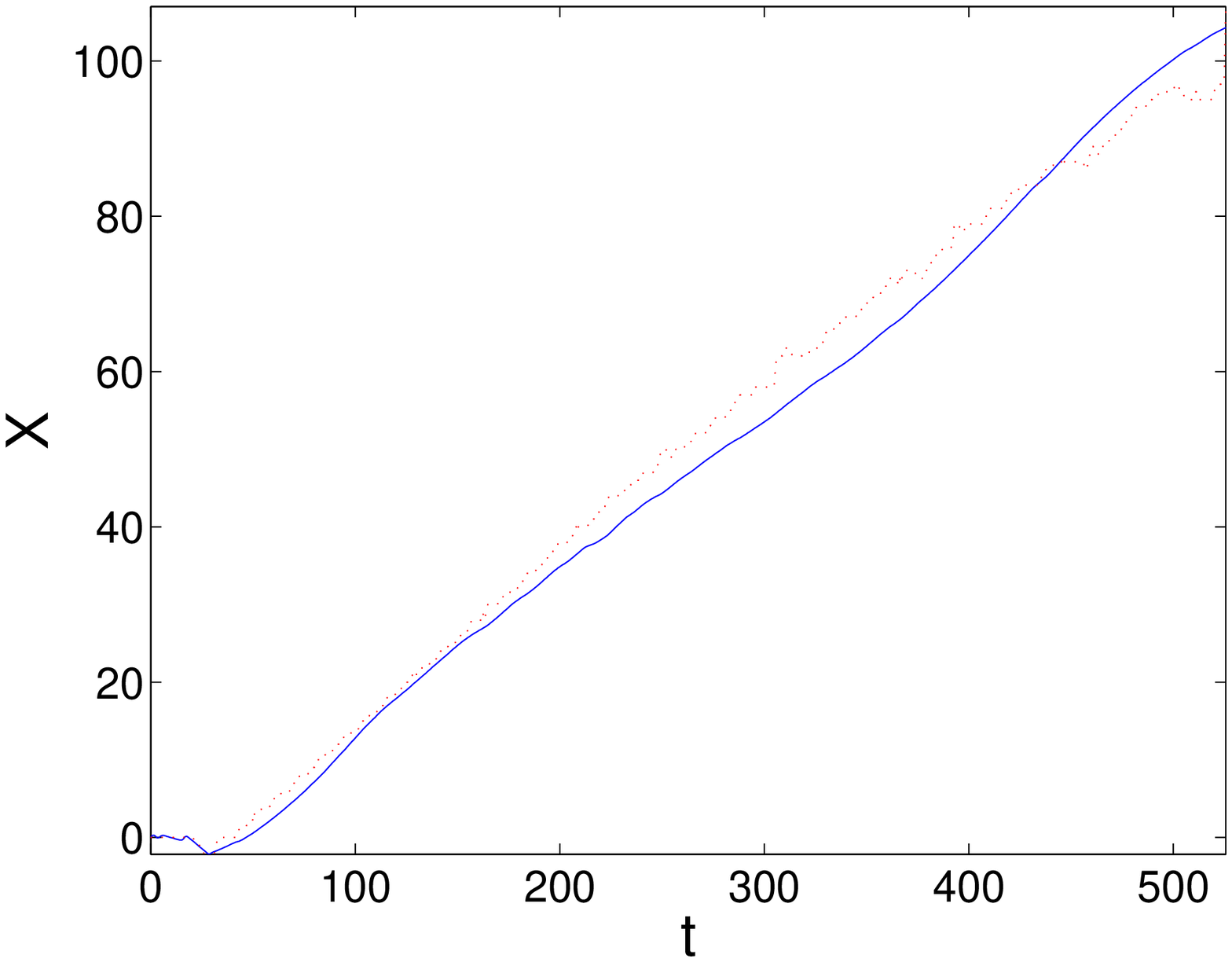} & \includegraphics[width=\middlefig]{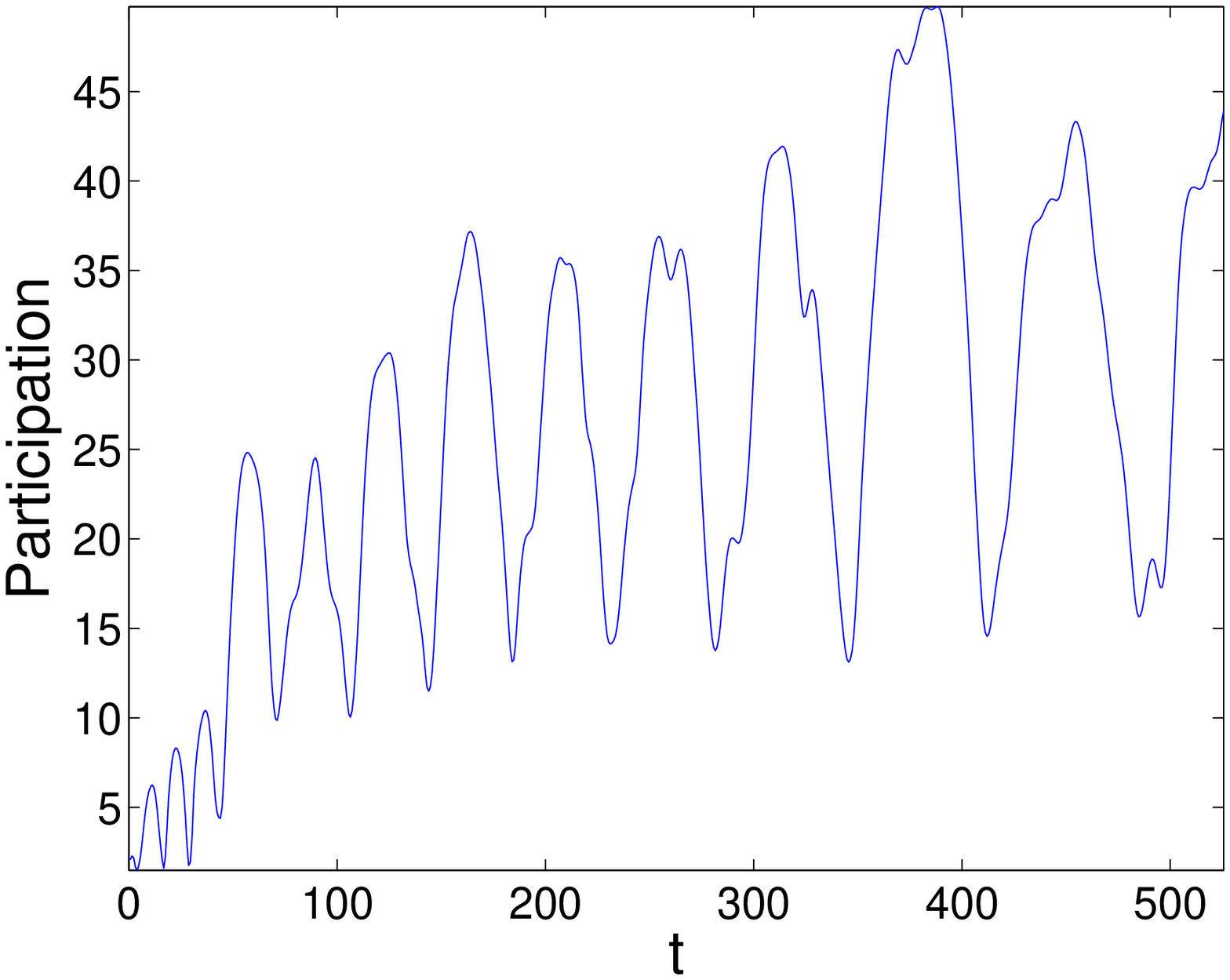} \\
(c) & (d) \\
\includegraphics[width=\middlefig]{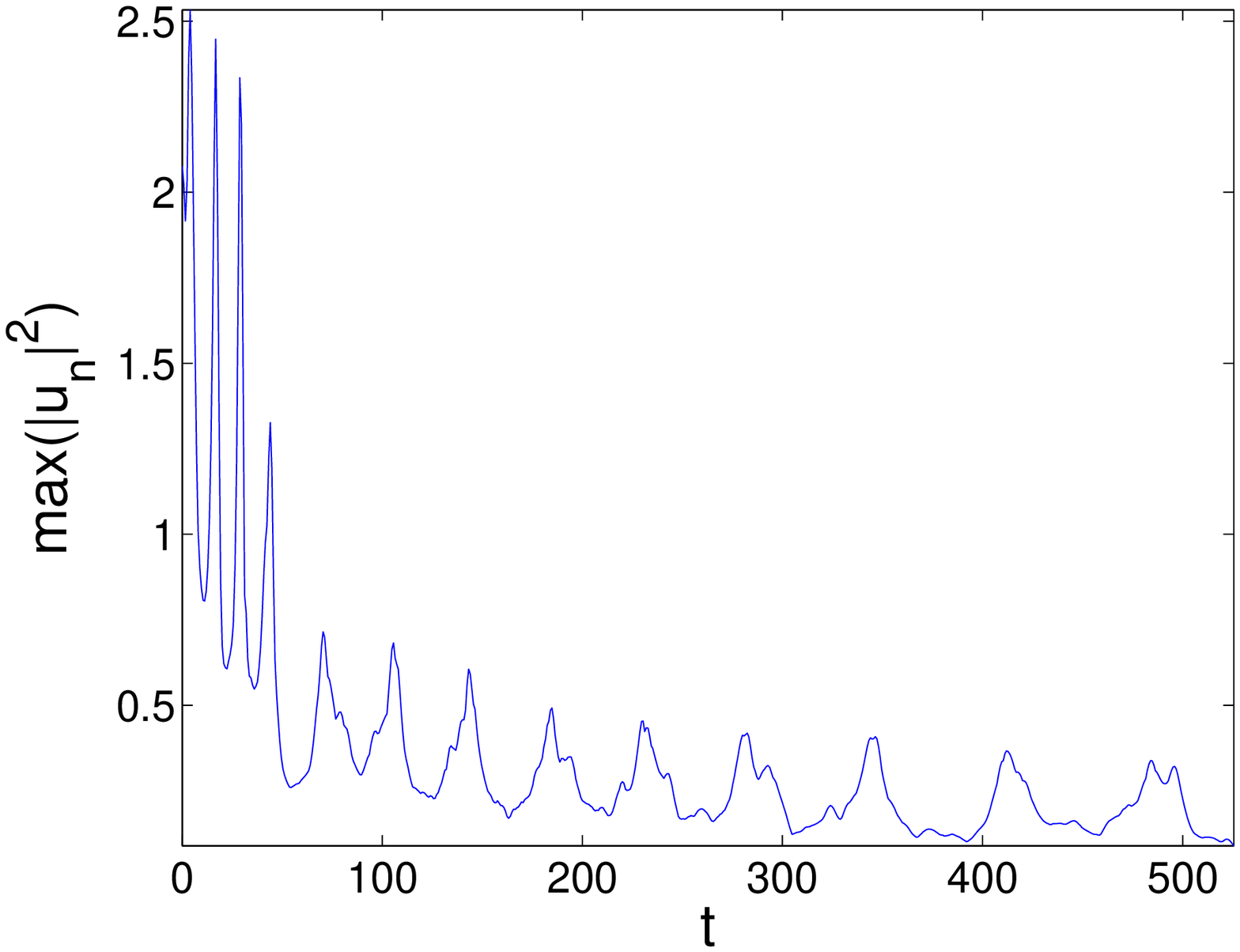} & \includegraphics[width=\middlefig]{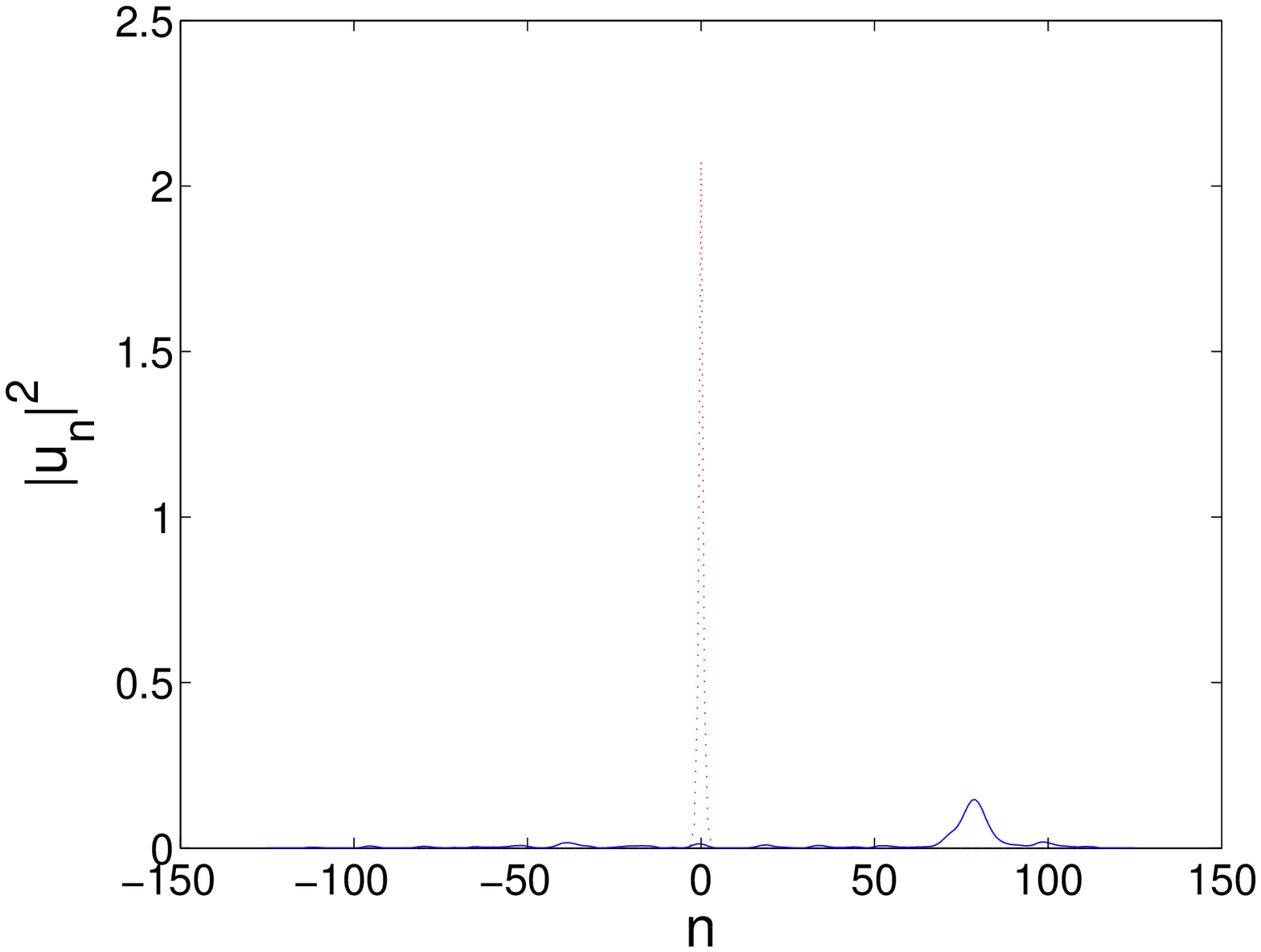}
\end{tabular}
\end{center}
\caption{A generic example of the free motion of a decaying soliton
in the straight direction, for $g_{\mathrm{ac}}=0.206$,
$\protect\omega =0.5$ and $\protect q =0.5$. In panel (d), the final
configuration pertains to $t_{\mathrm{fin}}=400$.} \label{fig4}
\end{figure}

\begin{figure}[tbp]
\begin{center}
\begin{tabular}{cc}
(a) & (b) \\
\includegraphics[width=\middlefig]{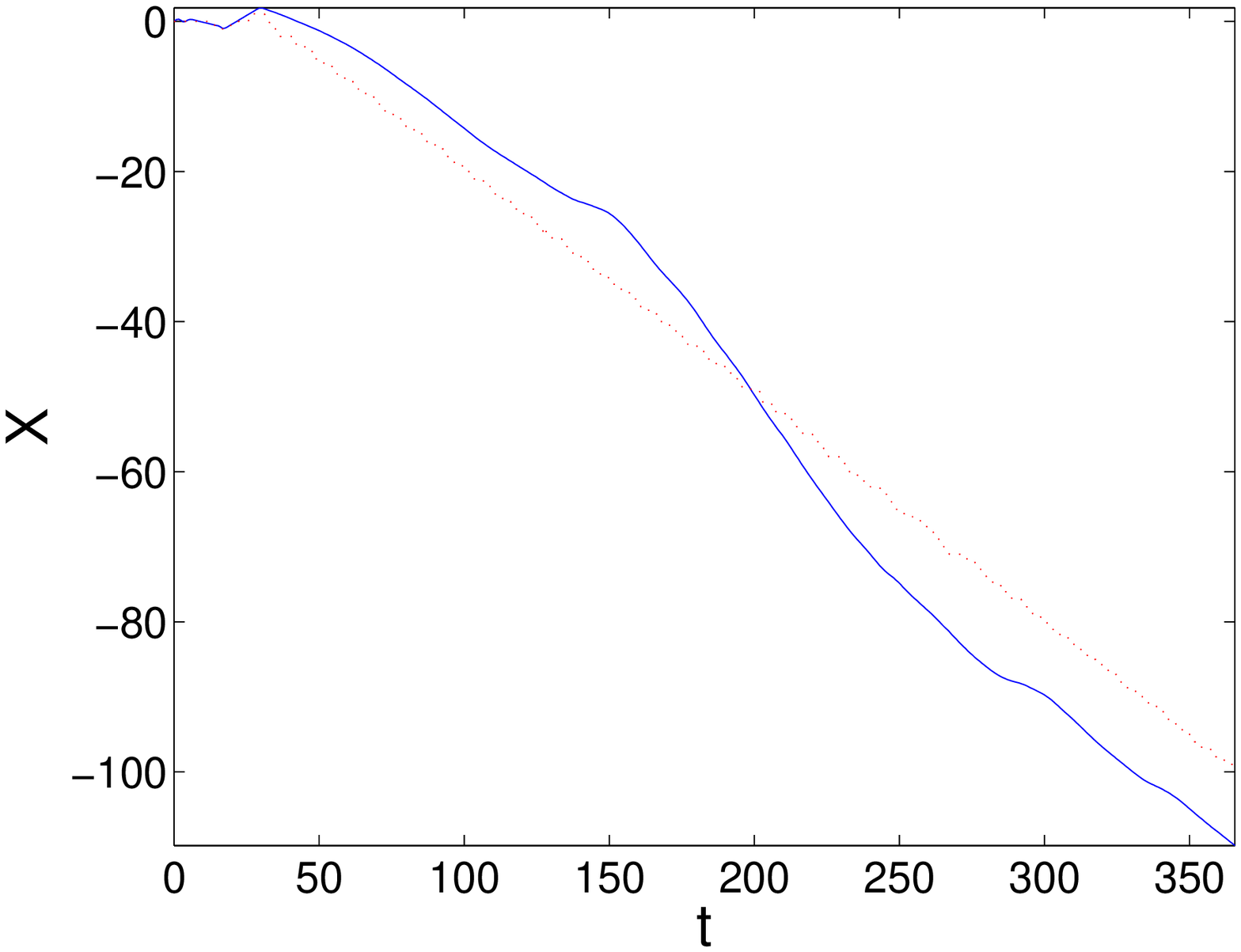} & \includegraphics[width=\middlefig]{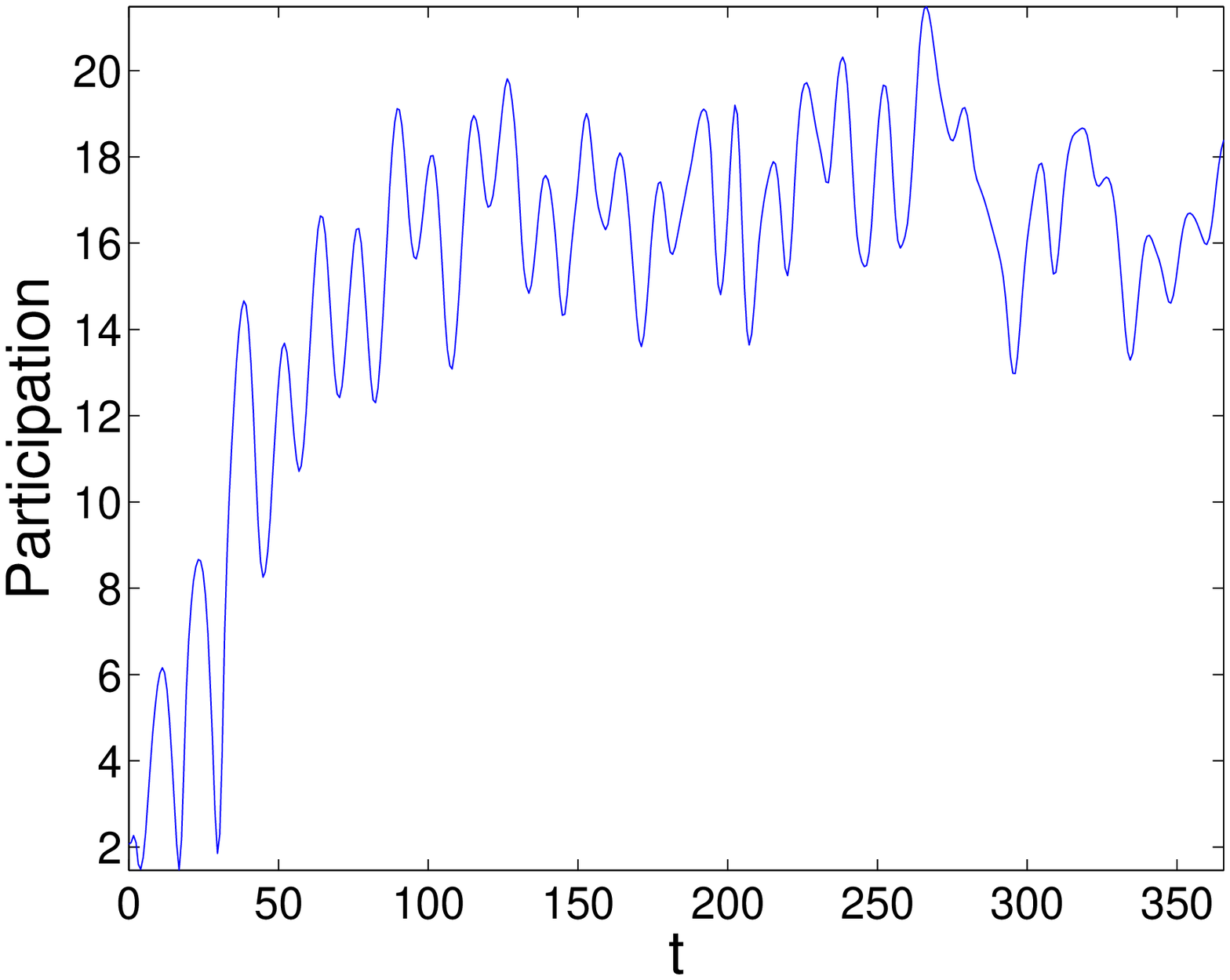} \\
(c) & (d) \\
\includegraphics[width=\middlefig]{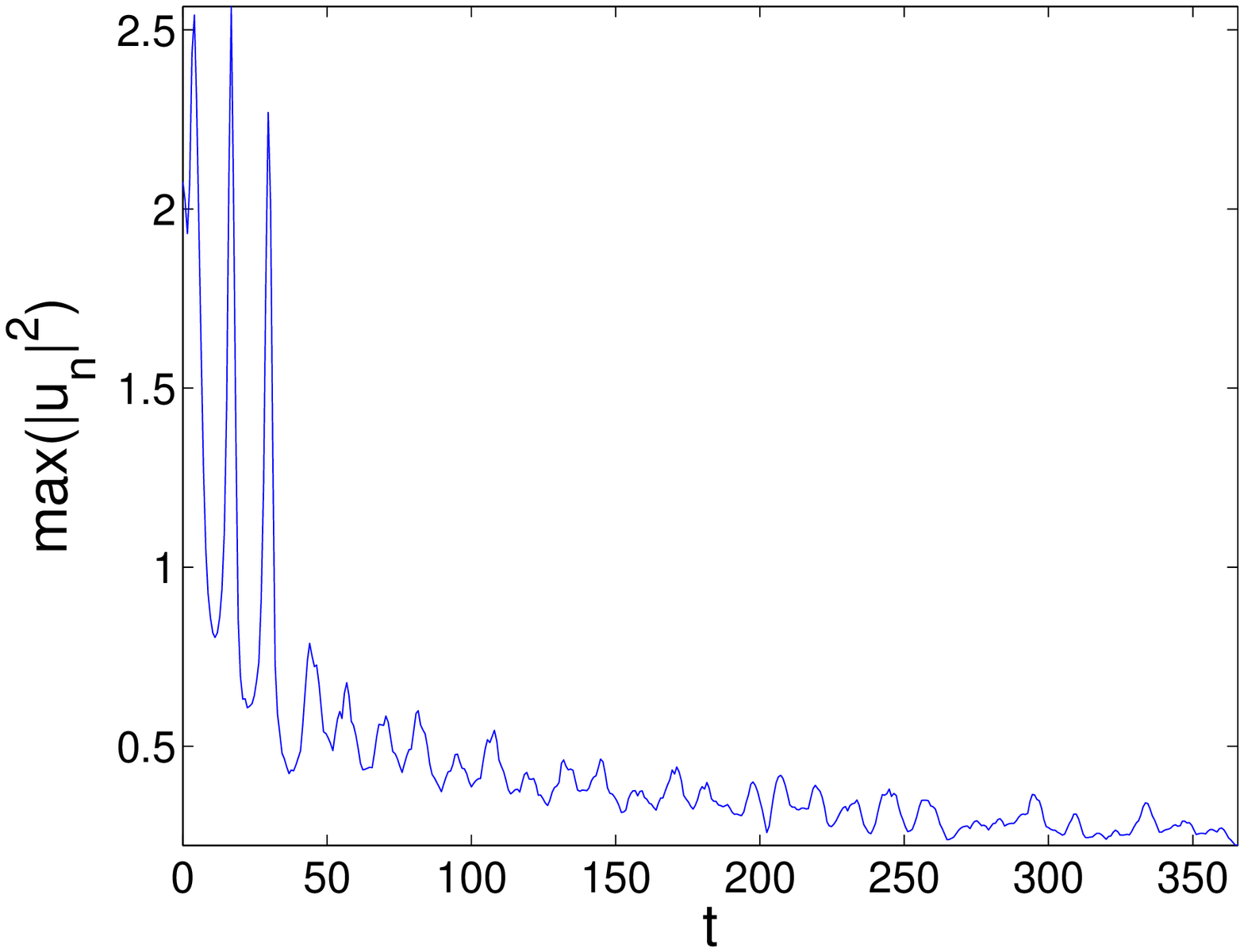} & \includegraphics[width=\middlefig]{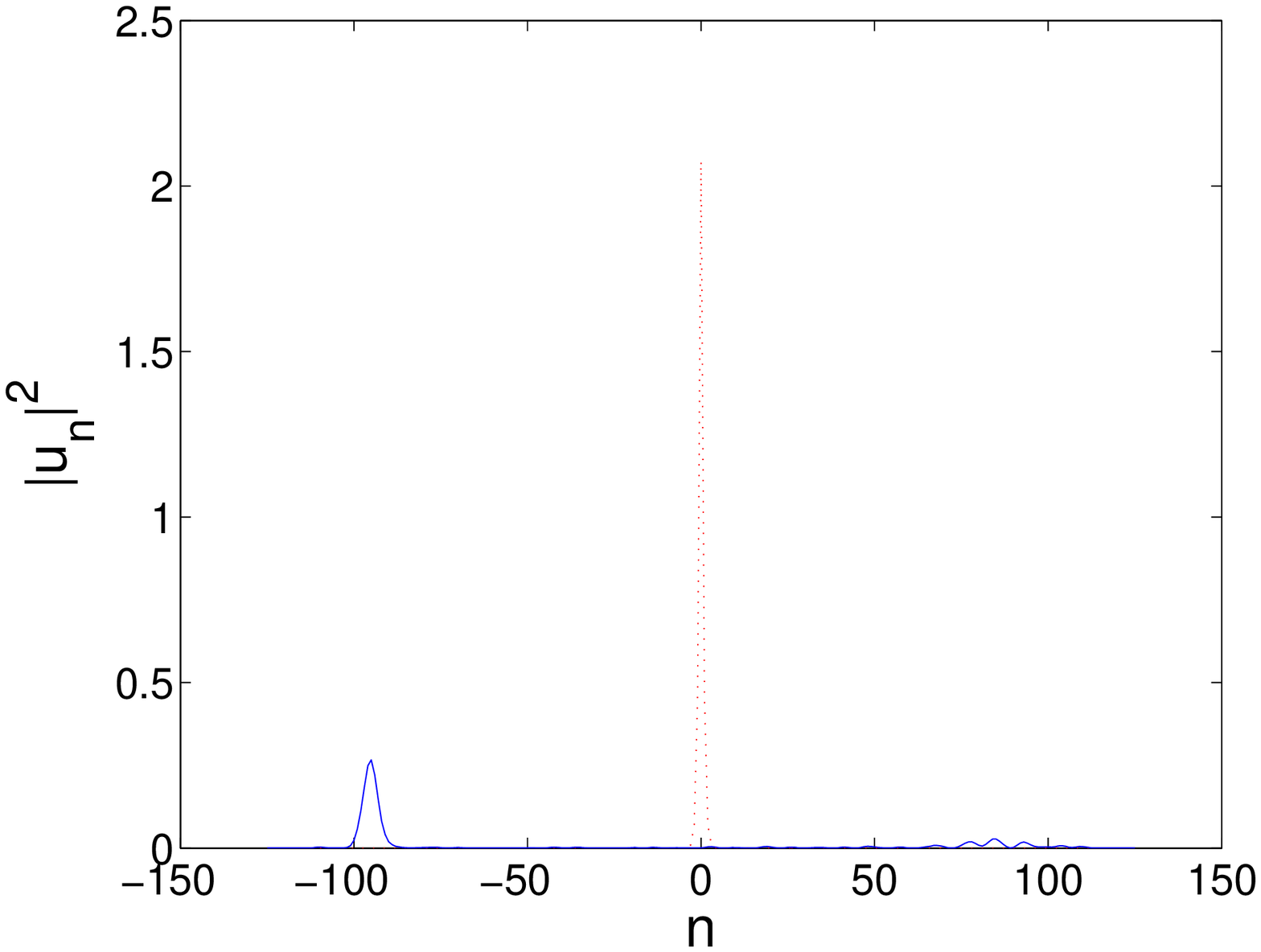}
\end{tabular}
\end{center}
\caption{The same as in Fig. \protect\ref{fig4}, in the case of
motion in the reverse direction, for $g_{\mathrm{ac}}=0.218$,
$\protect\omega =0.5$ and $\protect q =0.5$. In panel (d), the final
configuration is shown for $t_{\mathrm{fin}}=350$.} \label{fig5}
\end{figure}

\begin{figure}[tbp]
\begin{center}
\begin{tabular}{cc}
(a) & (b) \\
\includegraphics[width=\middlefig]{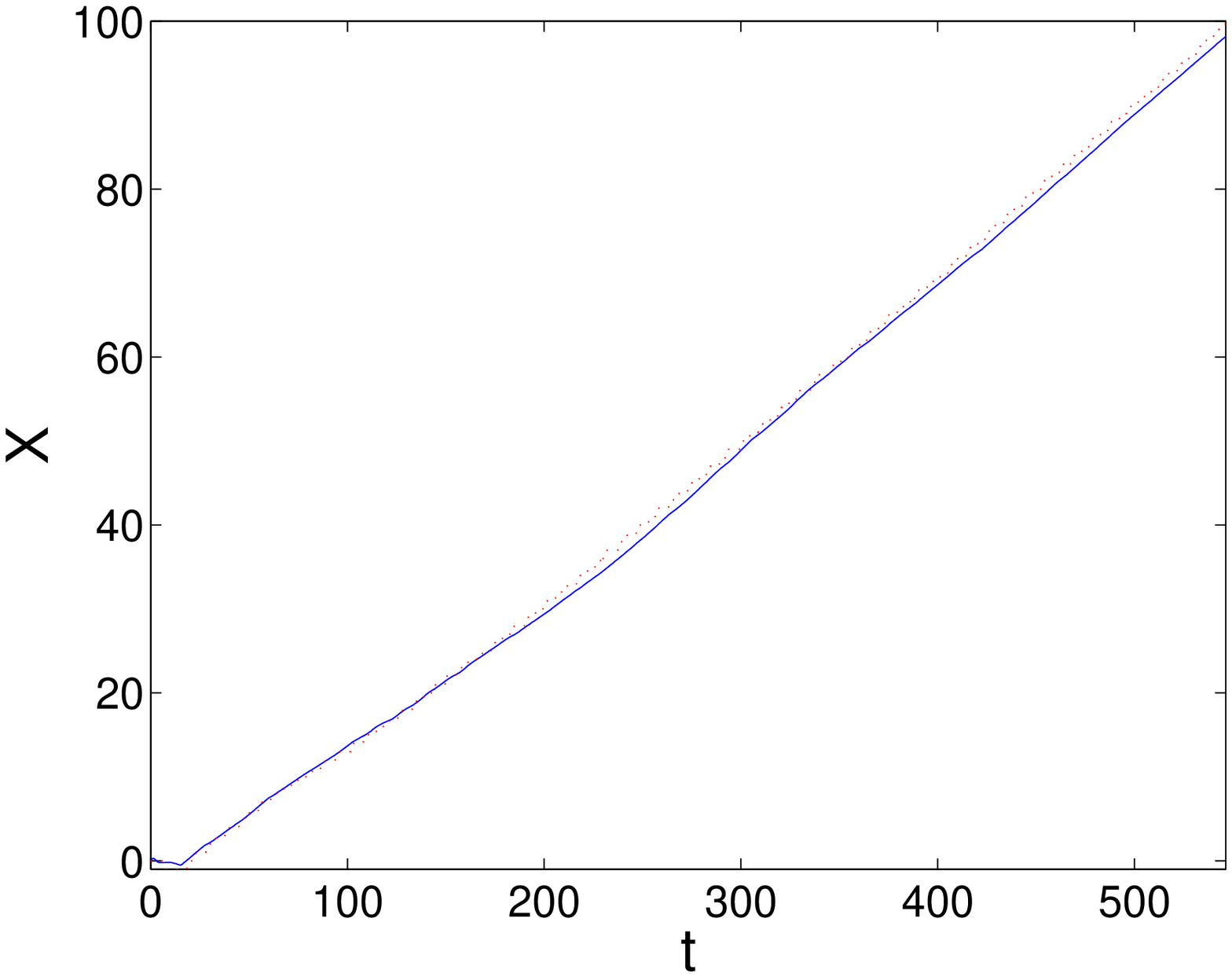} & \includegraphics[width=\middlefig]{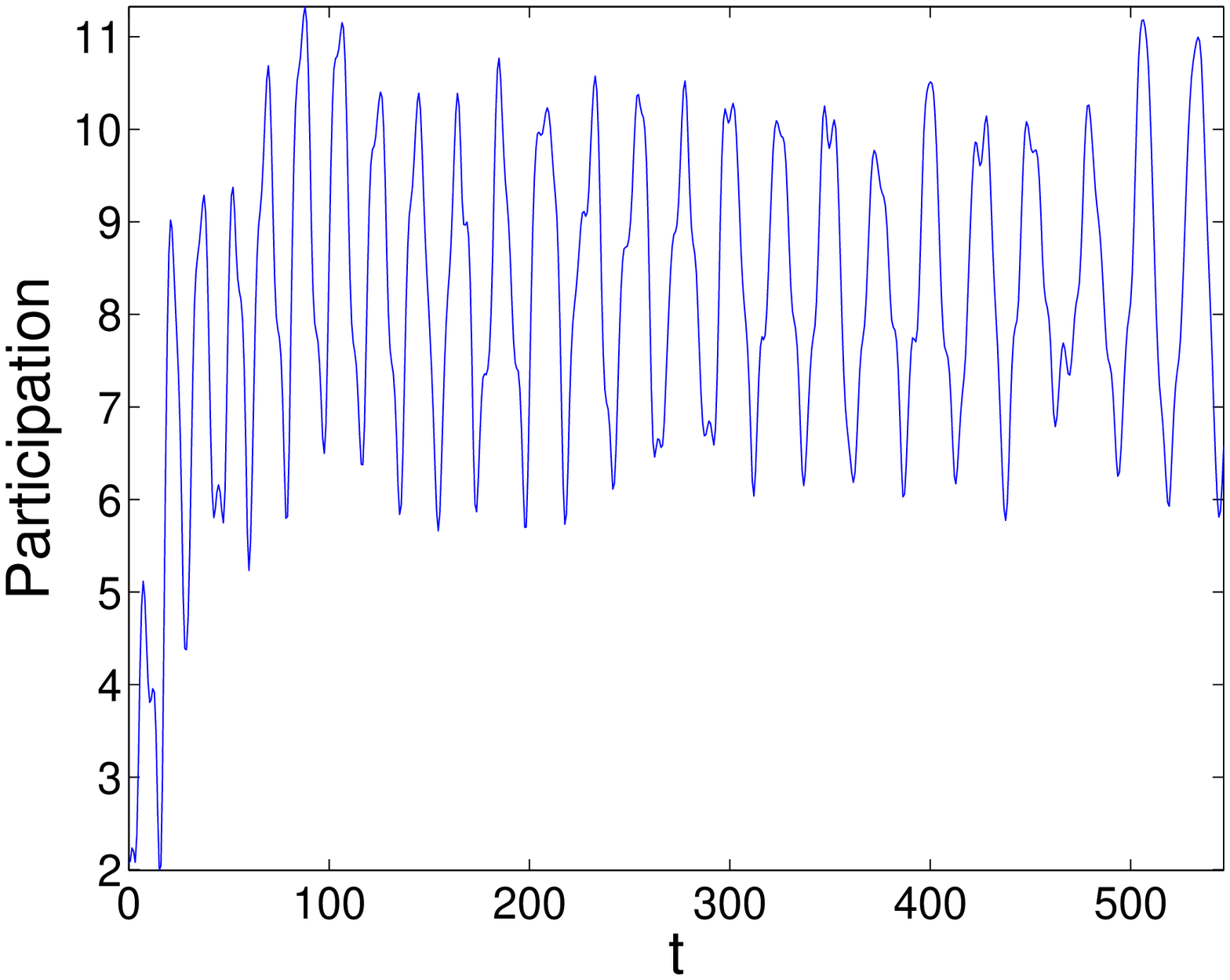} \\
(c) & (d) \\
\includegraphics[width=\middlefig]{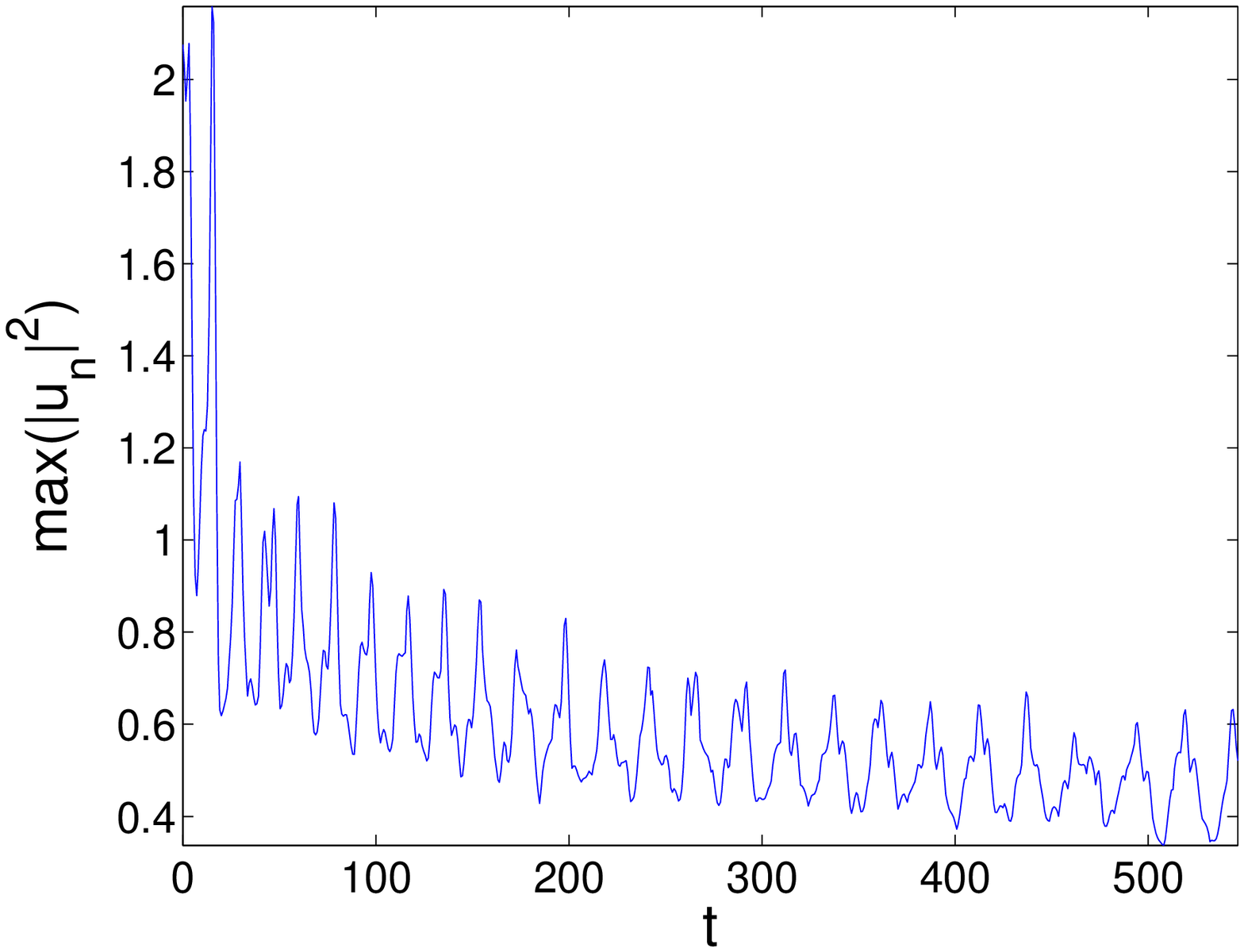} & \includegraphics[width=\middlefig]{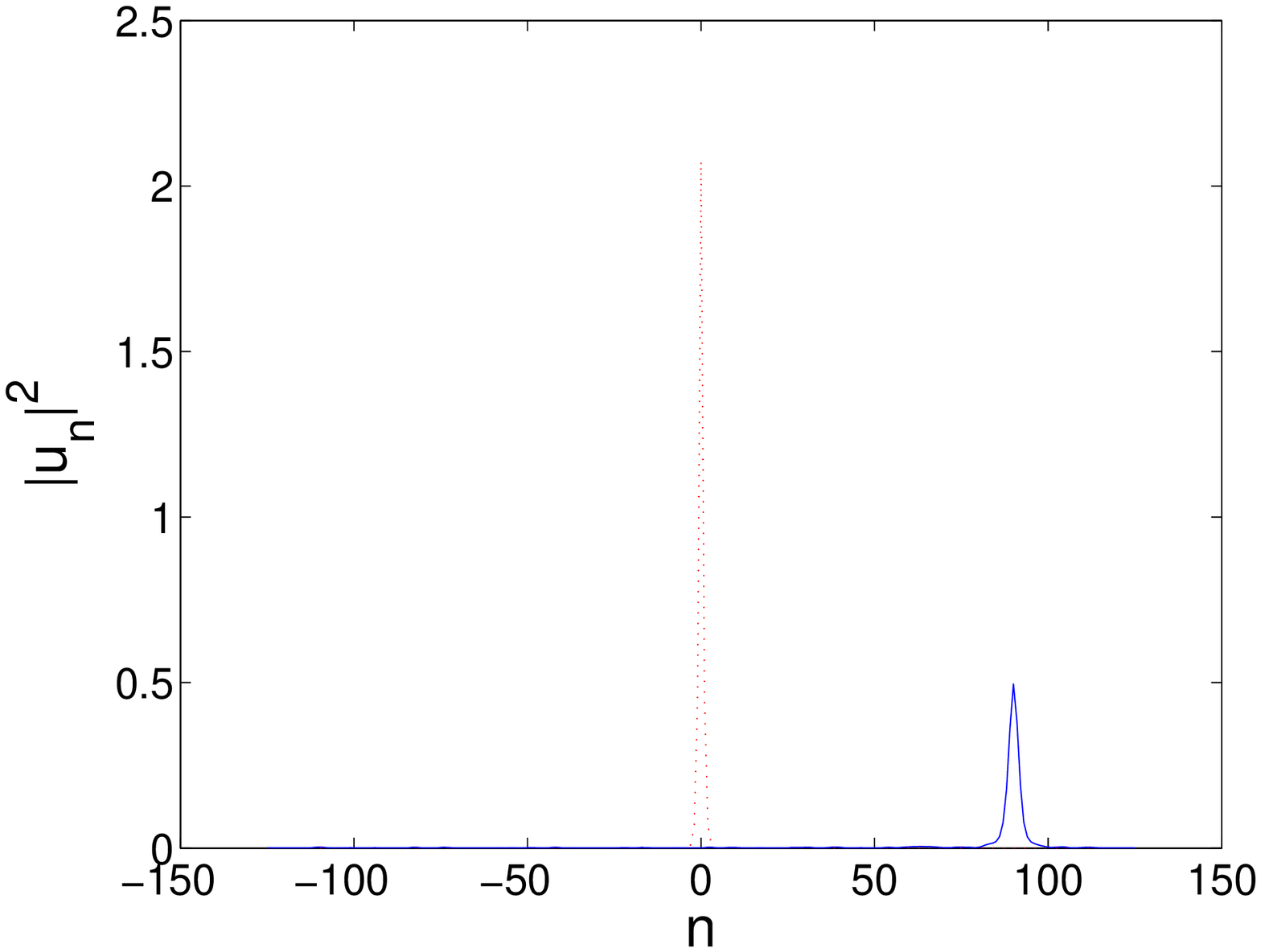}
\end{tabular}
\end{center}
\caption{A generic example of free motion of a \emph{stable}
(nondecaying) soliton in the straight direction, for
$g_{\mathrm{ac}}=0.132$, $\protect\omega =1$ and $\protect q =0.5$.
The panel (d) shows the final configuration at
$t_{\mathrm{fin}}=500$.} \label{fig6}
\end{figure}

\begin{figure}[tbp]
\begin{center}
\begin{tabular}{cc}
(a) & (b) \\
\includegraphics[width=\middlefig]{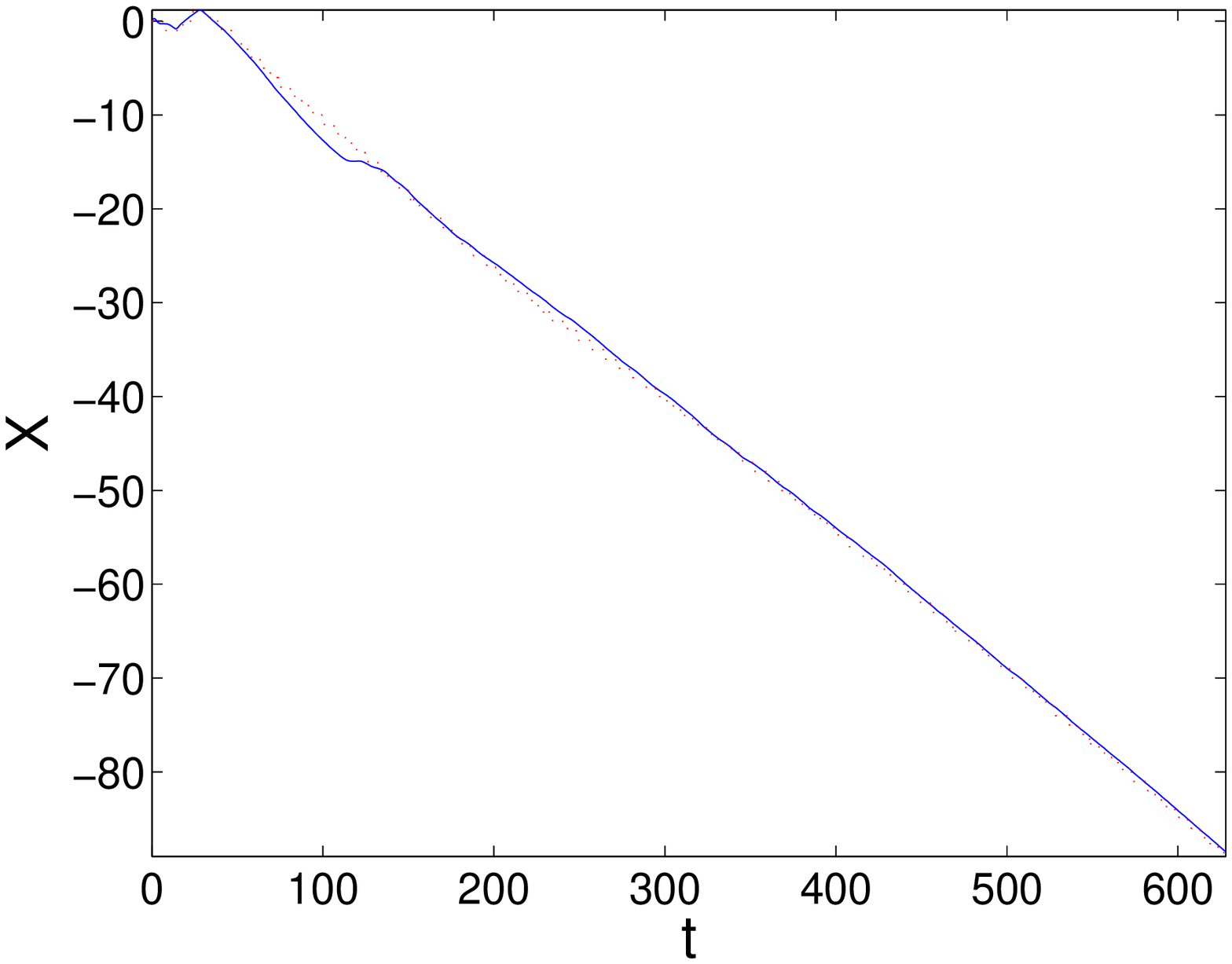} & \includegraphics[width=\middlefig]{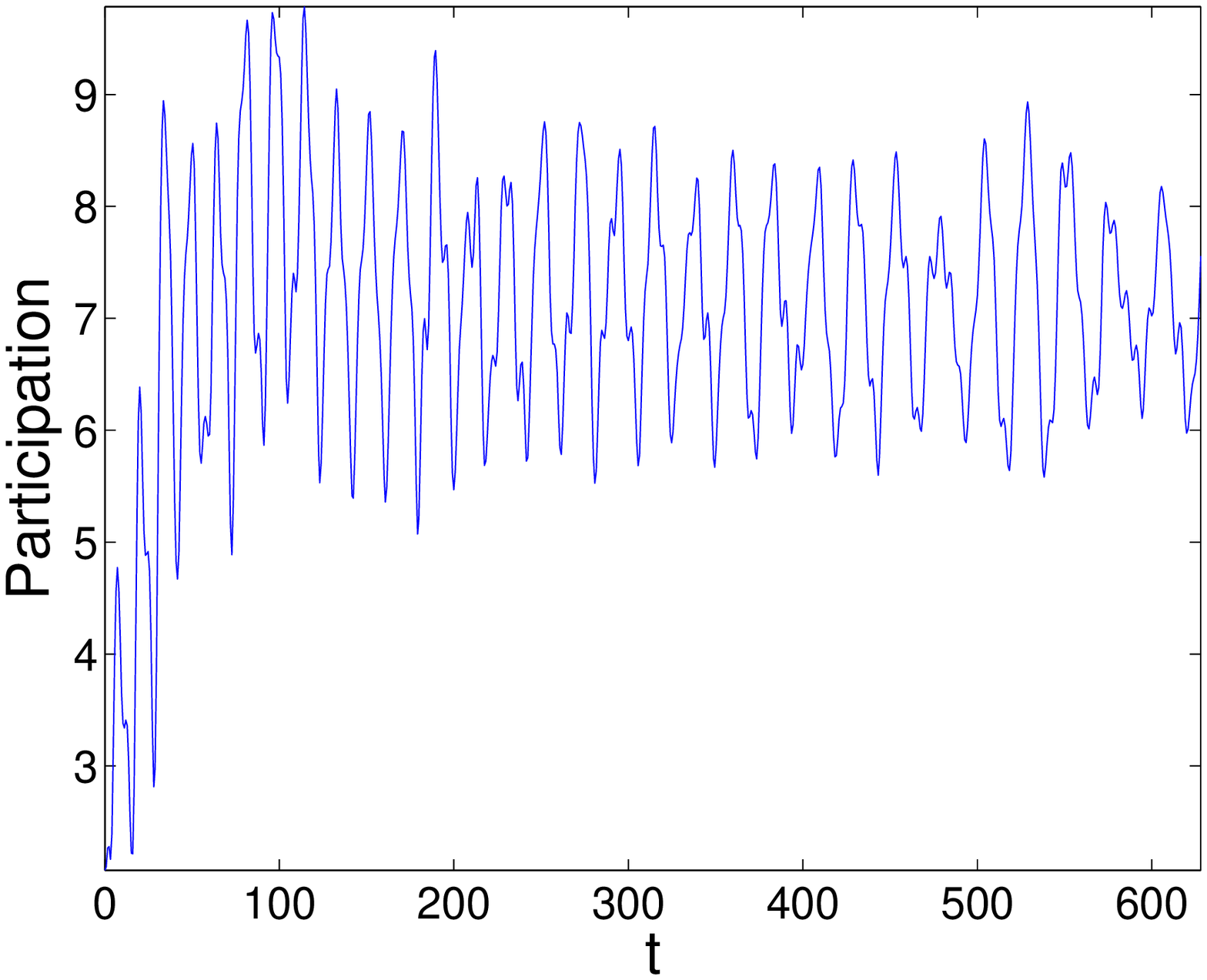} \\
(c) & (d) \\
\includegraphics[width=\middlefig]{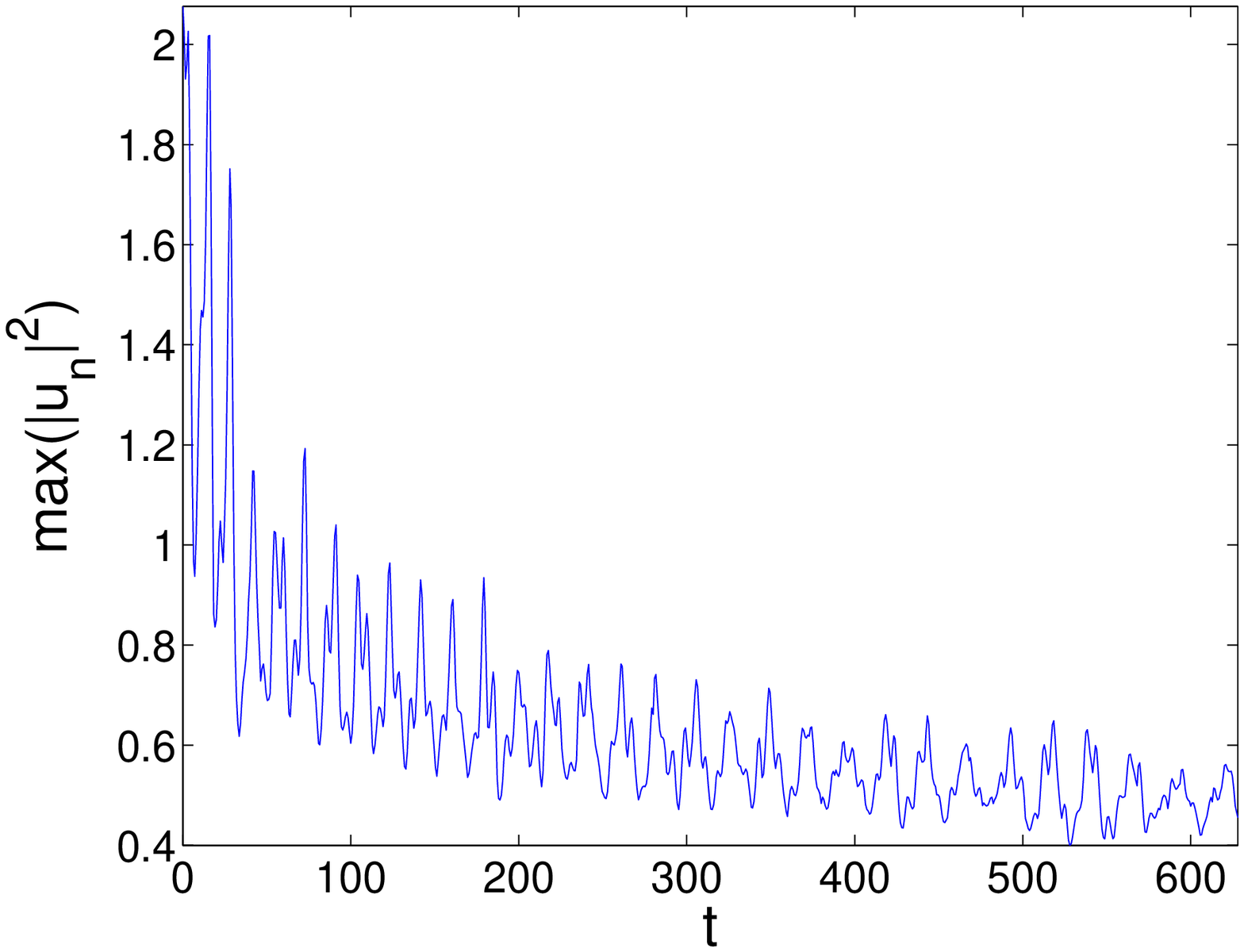} & \includegraphics[width=\middlefig]{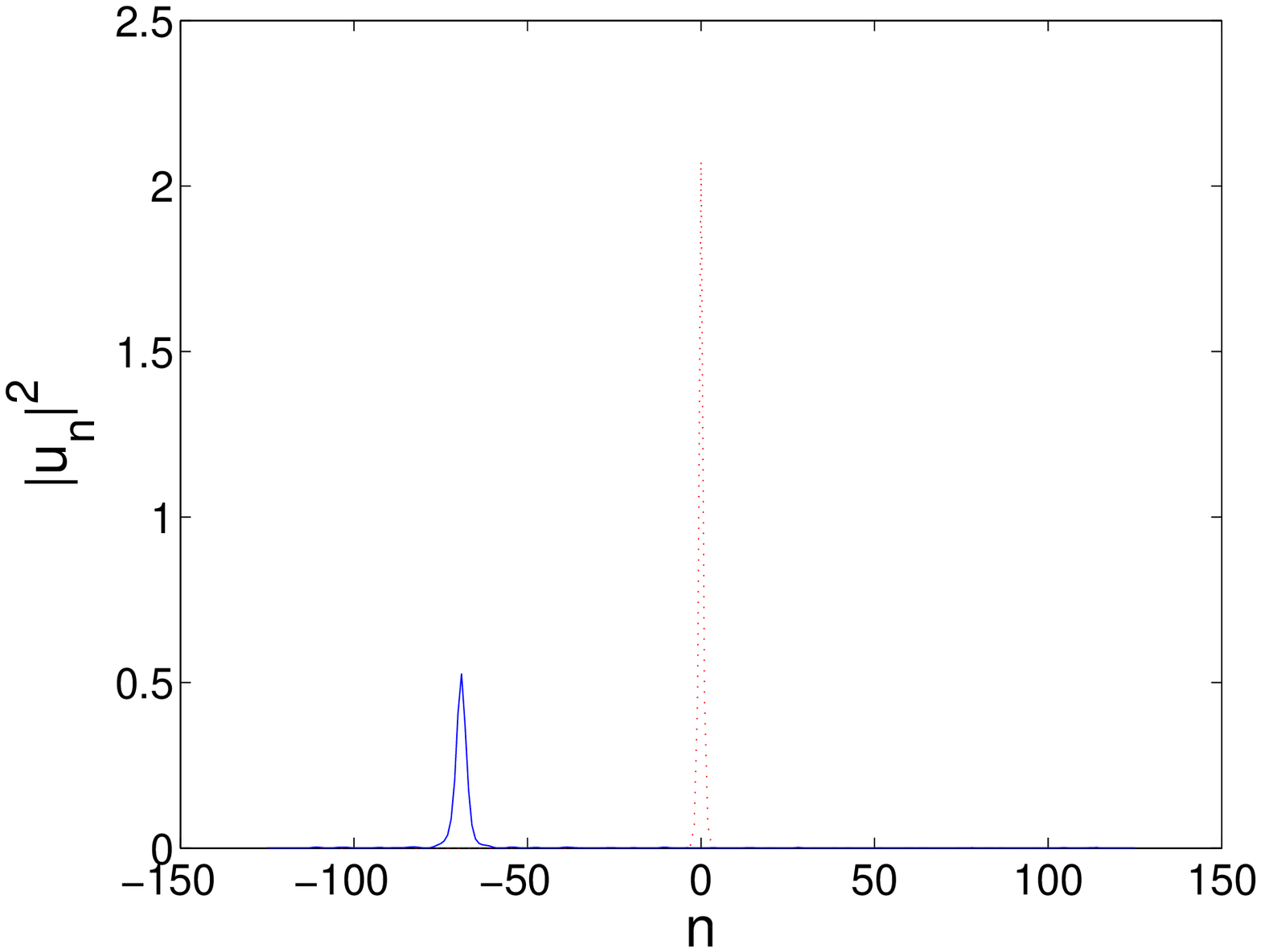}
\end{tabular}
\end{center}
\caption{The same as in Fig. \protect\ref{fig6}, but in the case of
motion of a stable soliton in the reverse direction, for
$g_{\mathrm{ac}}=0.122$, $\protect\omega =1$ and $\protect q =0.5$.
The panel (d) shows the final configuration at
$t_{\mathrm{fin}}=500$.} \label{fig7}
\end{figure}

Two distinct regimes of regular motion of the soliton were also
observed. In one case, shown in Figs. \ref{fig4} and \ref{fig5},
the moving soliton is not really stable, as it gradually decays
into radiation. The most interesting case is that of persistent
\emph{stable motion} of the soliton, without any observable decay
(after an initial transient stage of the evolution). Examples of
the latter are displayed in Figs. \ref{fig6} and \ref{fig7}. To
distinguish between the two different regimes of free motion, we
have adopted a criterion that the moving soliton is stable if it
keeps more than $70\%$ of the mass, $\sum \left\vert
u_{n}\right\vert ^{2}$, in its core.
While this criterion entails a degree of arbitrariness, we have
found that it accurately represents the dynamics of the structures
resulting from the FRM.
The generic examples
displayed in Figs. \ref{fig4} - \ref{fig7} demonstrate that free
motion of the soliton is possible in both the straight and reverse
directions, relative to the initial thrust.

\begin{figure}[tbp]
\begin{center}
\begin{tabular}{cc}
(a) & (b) \\
\includegraphics[width=\middlefig]{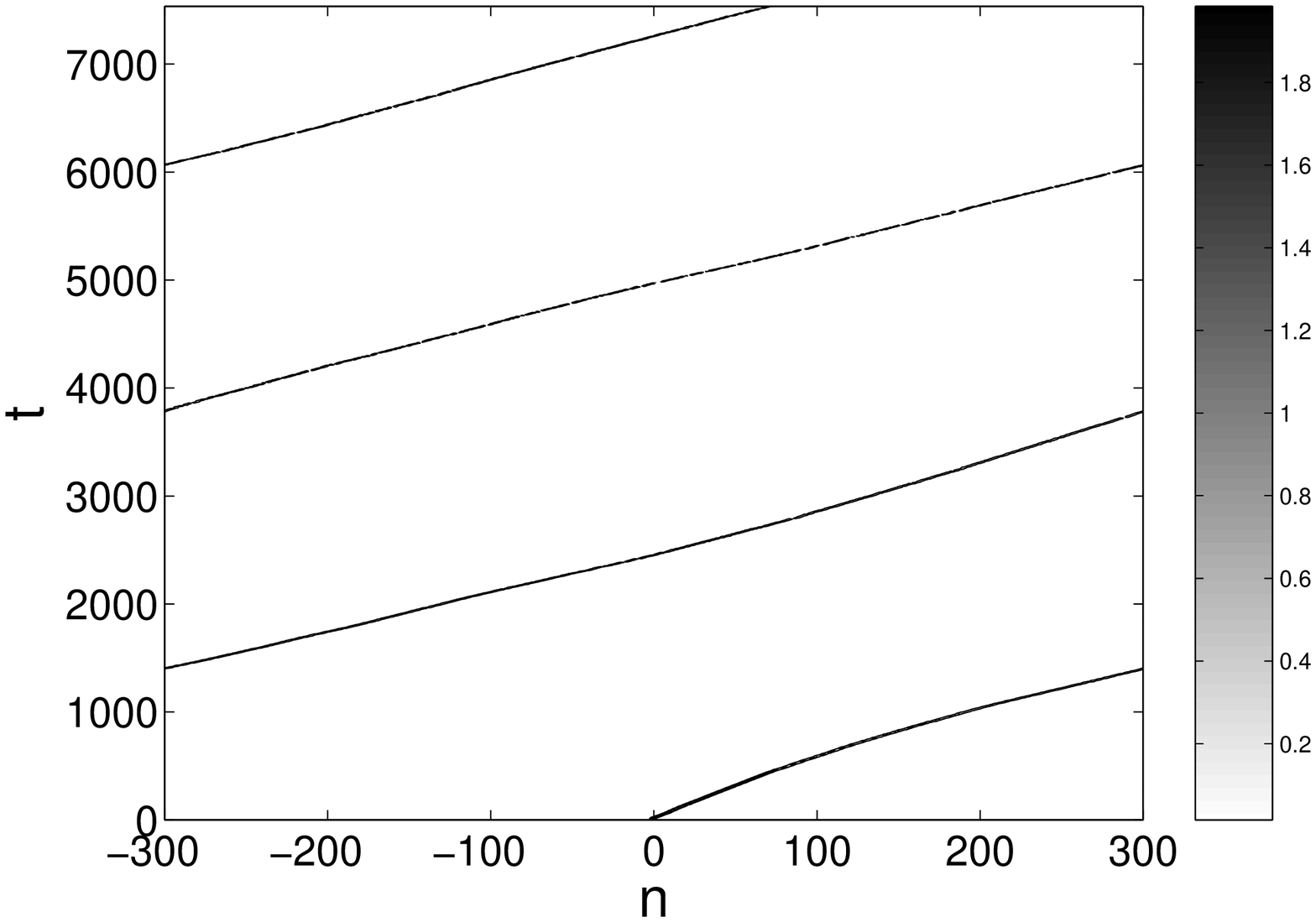} & \includegraphics[width=\middlefig]{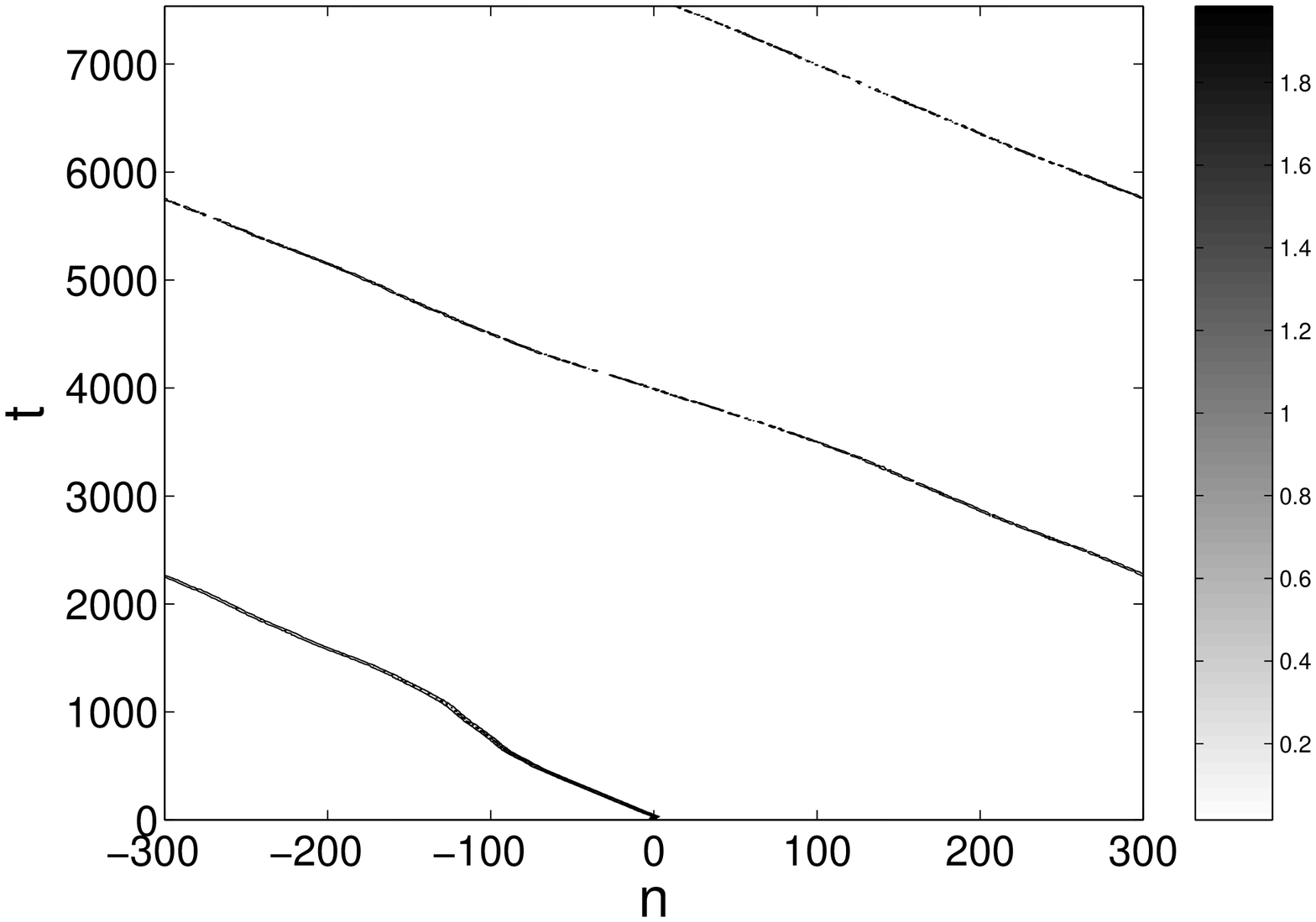}\end{tabular}\end{center}
\caption{Indefinitely long motion of stable solitons in a ring
lattice composed of $601$ sites. The gray-scale plots show the
spatiotemporal distribution of the density, $|u_{n}(t)|^{2}$. The
examples of the straight (a) and inverse (b) motion pertain,
respectively, to $g_{\mathrm{ac}}=0.132,\protect\omega =1, \protect
q =0.5$, and $g_{\mathrm{ac}}=0.122,\protect\omega =1,\protect q
=0.5$.} \label{fig8}
\end{figure}

As stable motion of solitons in the nonintegrable discrete model
is an issue of obvious interest, we have further investigated this
case, replacing the finite lattice with edge absorbers by a
ring-shaped one, with periodic boundary conditions. This setting
opens a way to study indefinitely long motion of the soliton. The
result, illustrated by examples shown in Fig. \ref{fig8}, is that
the moving solitons remains stable (preserving its shape) as long
as the simulations could be run. In this case, it is natural to
compare the average velocity $\bar{c}$ of the persistent motion
with the prediction given by Eq. (\ref{res}). The result is
$\bar{c}_{1}\approx 0.246$ and $\bar{c}_{2}\approx 0.155$ in the
cases shown in Figs. \ref{fig8}(a) and and Fig. \ref{fig8}(b),
respectively. Comparison with the analytical formula (\ref{res})
(with $\omega =1$, which is the driving frequency in the examples
shown in Fig. \ref{fig8}) demonstrate that $\bar{c}_{2}$ and
$\bar{c}_{1}$ fit well to the predicted values in the cases of
the, respectively, fundamental and second-order resonance,
\begin{equation}
\bar{c}_{2}/\left( c_{\mathrm{res}}\right) _{1}^{(1)}\approx
0.974,\bar{c}_{1}/\left( c_{\mathrm{res}}\right) _{3}^{(2)}\approx
1.029 \label{cc}
\end{equation}Relatively small discrepancies between the predicted and observed values in
Eq. (\ref{cc}) can be explained by the fact that the effective perturbations
are not really weak in these cases.

\begin{figure}[tbp]
\begin{center}
\begin{tabular}{cc}
\multicolumn{2}{c}{$q=0.25$} \\
\includegraphics[width=\middlefig]{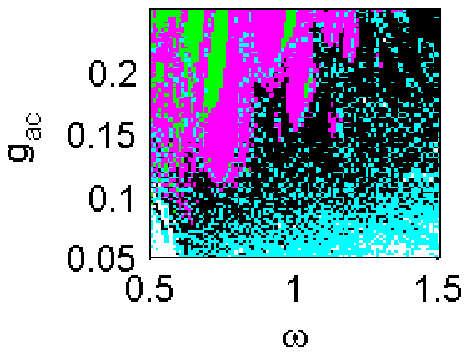} & \includegraphics[width=\middlefig]{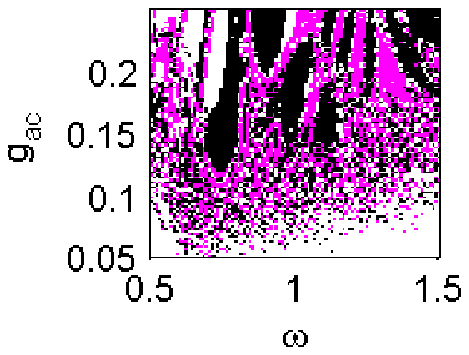} \\
\multicolumn{2}{c}{$q=0.5$} \\
\includegraphics[width=\middlefig]{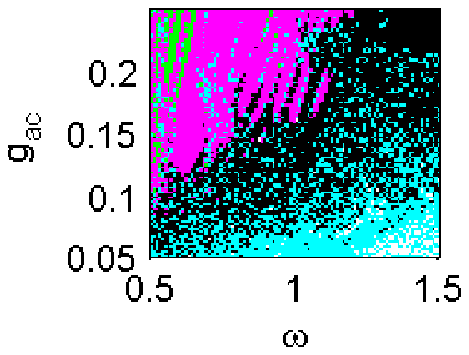} & \includegraphics[width=\middlefig]{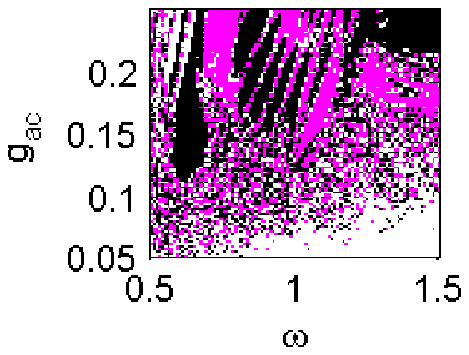} \\
\multicolumn{2}{c}{$q=1$} \\
\includegraphics[width=\middlefig]{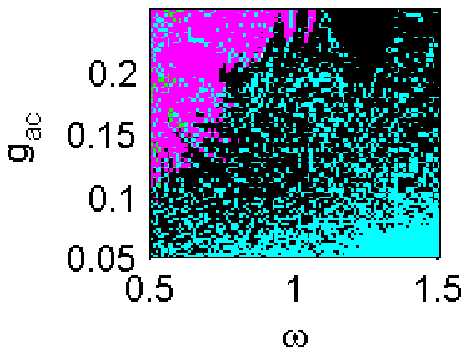} & \includegraphics[width=\middlefig]{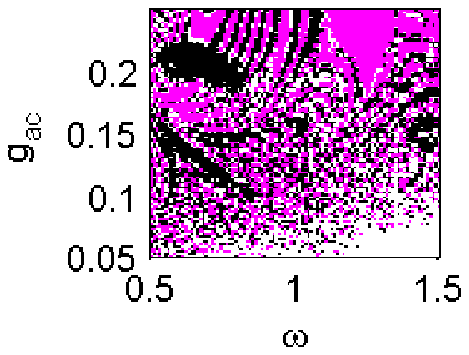} \\
&
\end{tabular}\end{center}
\caption{Maps in the left column show areas in the parameter plane
$(\protect\omega ,g_{\mathrm{ac}})$ which give rise to the
following dynamical regimes. White areas: the soliton remains
pinned; cyan: irregular motion; green: splitting; magenta:
regular motion with decay; black: stable motion (without decay).
The maps in the right column additionally show the difference
between the forward (alias straight, marked by magenta) and
backward (alias reverse, marked by black) directions of the
regular motion, relative to the direction singled out by the
initial push. Regular-motion regimes for both decaying and stable
solitons are included here.} \label{fig9}
\end{figure}

%\textit{Global description of the dynamical regimes }
For the fixed values of
$g_{\mathrm{dc}}=1$ in Eq. (\ref{g}) and $\nu =1$ in Eq. (\ref{initial}),
and several values of the initial thrust in Eq. (initial), $q=0.25$, $0.5$,
and $1$, we have collected results of systematic simulations, varying the
drive's parameters, $g_{\mathrm{ac}}$ and $\omega $, by small steps in broad
ranges spanning the relevant two-parameter space.
The results are encompassed in Fig. \ref{fig9}, in the form of maps in
the $\left( \omega ,g_{\mathrm{ac}}\right) $ plane, where we outline regions
giving rise to each of the qualitatively different dynamical regimes
described above. This road map is one of the key findings of the
present work, detailing the various possibilities arising as a result
of FRM in a dynamical lattice.

Some general features can be deduced from examination of the maps
in Fig. \ref{fig9}. As is seen, the increase of the initial thrust
$q$ significantly affects the map, although quantitatively, rather
than qualitatively. At all values of $q$, the irregular motion is,
generally, changed by stable progressive motion (straight or
reverse) with the increase of the drive's amplitude, and/or
decrease of its frequency, which seems quite natural. Further
increase of the drive's strength, which implies the applications
of a relatively strong perturbation to the system, may be expected
to lead to an instability, which indeed happens, in the form of an
onset of the gradual decay of the moving solitons. Finally, strong
instability sets in, manifesting itself in the splitting of the
soliton. It also seems natural that the soliton is more prone to
splitting if the driving frequency is low, as internal strain in
the pulse, which eventually leads to its splitting, has more time
to accumulate if the drive oscillates slowly.

Transition to the reverse motion tends to happen parallel to the transition
from the stable moving soliton to the decaying one. For this reason, in most
cases (but not always) backward-moving soliton are decaying ones. Finally,
it should be noted that the increase of the initial thrust leads to
overall \emph{stabilization} of the soliton (somewhat counter-intuitively),
making the decay and splitting zones smaller.

\begin{figure}[tbp]
\begin{center}
\begin{tabular}{cc}
(a) & (b) \\
\includegraphics[width=\middlefig]{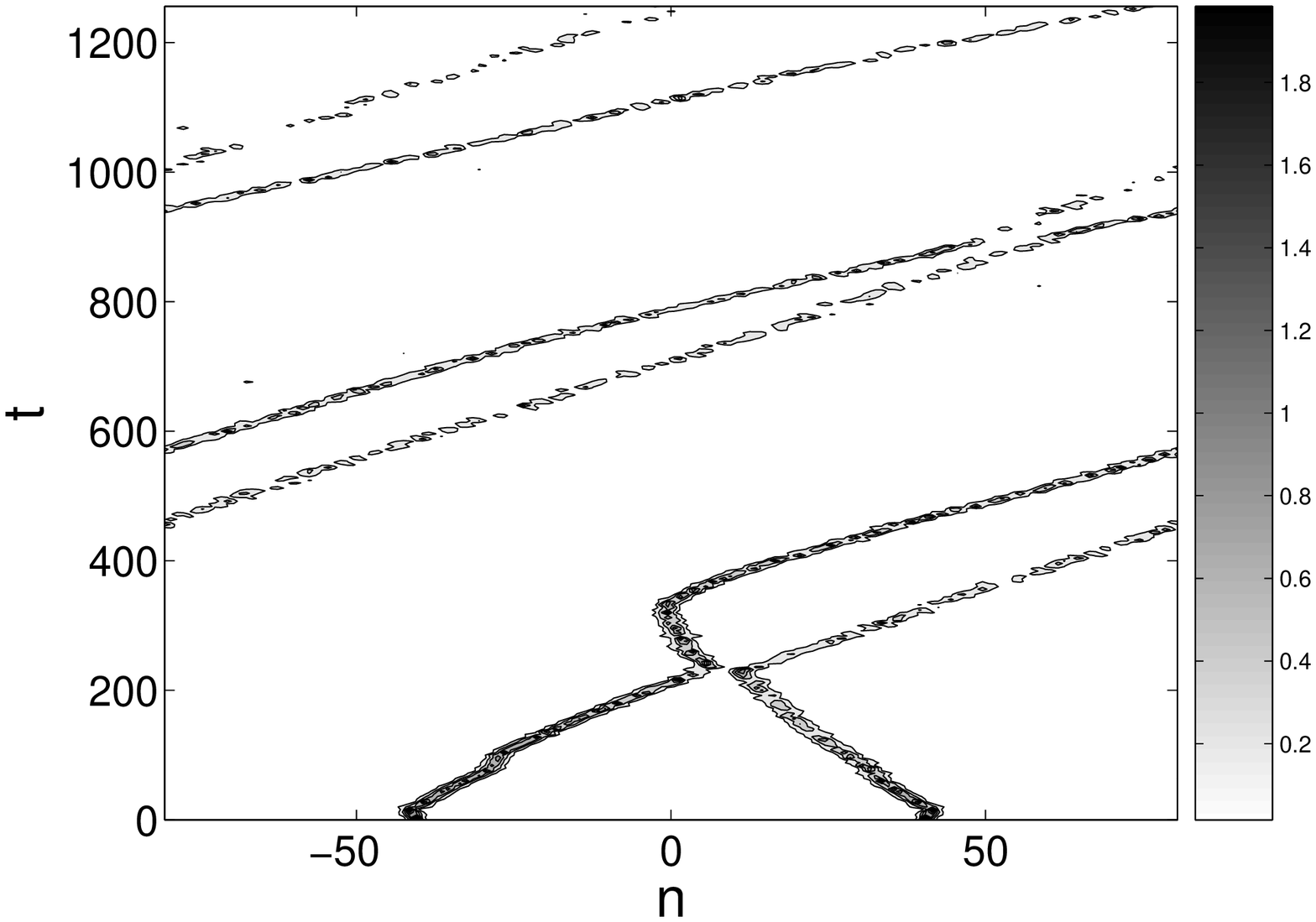} & \includegraphics[width=\middlefig]{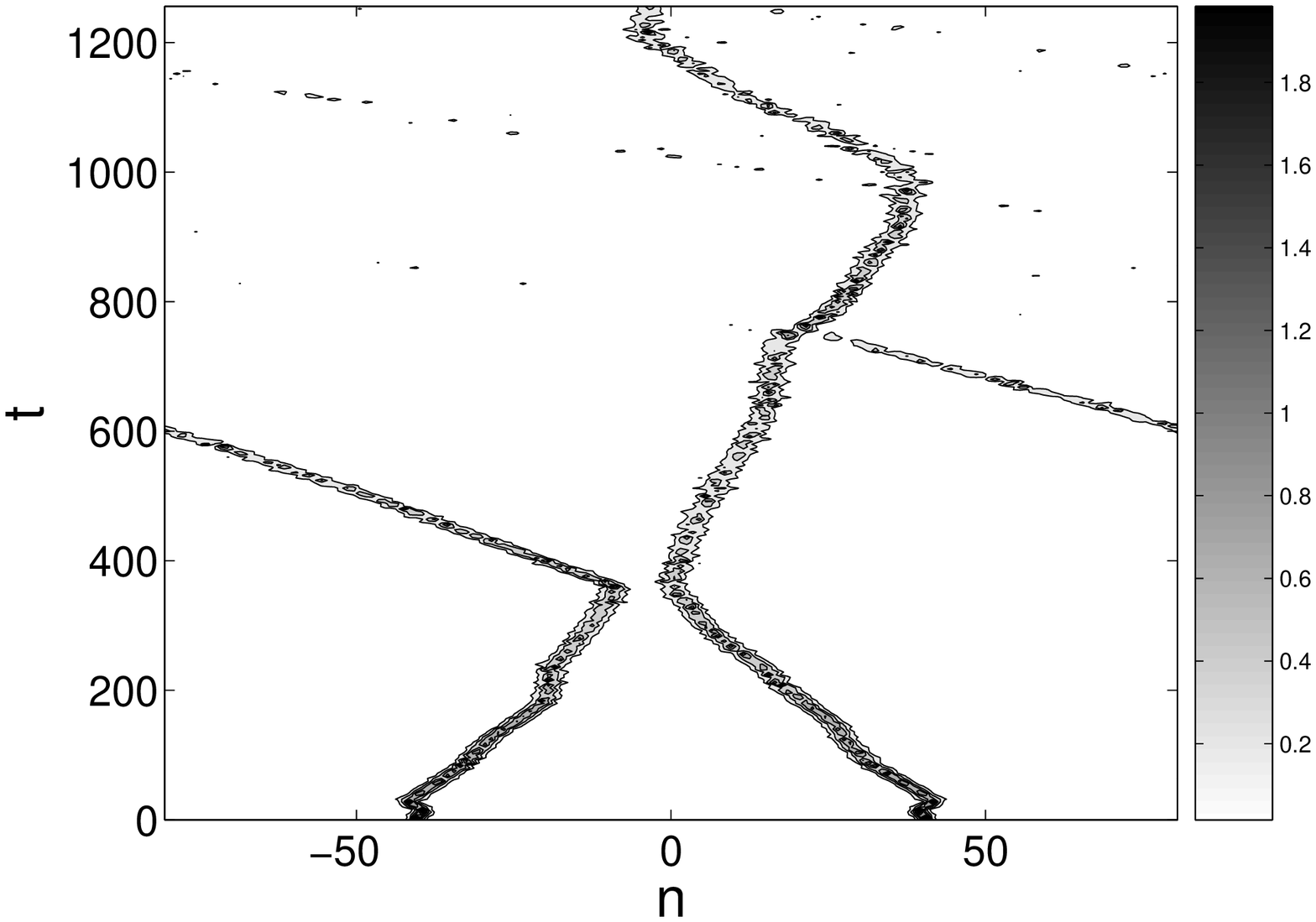} \\
&
\end{tabular}\end{center}
\caption{Two typical examples of different outcomes of collisions
between solitons with equal masses moving in opposite directions in
the lattice with periodic boundary conditions. The parameters are
$g_{\mathrm{ac}}=0.132$, $\protect\omega =1$, $\protect q =0.5$ (a)
and $g_{\mathrm{ac}}=0.122$, $\protect\omega =1$, $\protect q =0.5$
(b). Note that the collision is multiple in panel (b).}
\label{fig10}
\end{figure}

%\textit{Collisions between solitons}
Finally, using the large lattice with the
periodic boundary conditions, we also simulated collisions between solitons
originally moving with opposite velocities. Initial pulses were generated by
applying the thrust $\pm q$ to two quiescent solitons. A systematic study of
collisions is very difficult in the present model, cf. Ref. \cite{Amazon}.
Nevertheless, we were able to identify two different types of the
interaction, typical examples of which are shown in in Fig. \ref{fig10}.

In the case of Fig. \ref{fig10}(a), the solitons bounce back from each other
quite elastically. Afterwards, one of the soliton spontaneously reverses its
direction of motion, due to its interaction with the underlying lattice.
Eventually, we observe a pair of virtually noninteracting solitons traveling
indefinitely long in the same direction.

In another case, Fig. \ref{fig10}(b), the solitons also bounce after the
first collision; however, in this case the collisions are inelastic,
resulting in transfer of mass from one soliton to the other. Repeated
collisions lead to additional transfer, and eventually the weak soliton
almost disappears.
%It should be noted that, in contrast to what is known
%about collisions between moving solitons in the ordinary DNLS equation (the
%one with constant coefficients) \cite{Amazon}, we have never observed merger
%of colliding solitons into a standing one.

%\textit{Conclusion}

\section{Coclusions}

In this work, we have investigated moving solitons in
the DNLS equation with a periodically time-modulated nonlinear coefficient.
An approximate analytical consideration predicts that the ac nonlinearity
management may facilitate creation of traveling solitons in the lattice.
Systematic simulations reveal several generic dynamical regimes, depending
on parameters of the time modulation, and the initial thrust which sets the
soliton in motion. Besides the possibility that the soliton remains pinned,
the basic dynamical regimes feature irregular motion, regular motion of a
decaying soliton, and regular motion of a stable one. In the latter case,
extremely long simulations in the lattice with periodic boundary conditions
demonstrate that the soliton keeps moving indefinitely long without any
tangible loss. Velocities of the moving stable solitons are found to be in
good agreement with the analytical prediction
through a resonance condition, revealing the key mechanism for
sustaining the travelling motion.
%The results reported in this
%work actually offer the first example of the resonantly ac-driven
%progressive motion of a \emph{nontopological} soliton.
All the generic
dynamical regimes were mapped in the model's parameter space. Collisions
between stable moving solitons were briefly investigated too, with a
conclusion that two different outcomes are possible: elastic bounce, and
bounce with mass transfer from one soliton to the other. The model can be
realized in a Bose-Einstein condensate trapped in a deep optical lattice,
with the nonlinearity modulation induced by the Feshbach resonance in the ac
regime.

It would be interesting to examine similar dynamical scenaria for
the effect of FRM on dark solitons, and also to monitor how
this phenomenology is modified in higher dimensions. Such studies
are currently in progress and will be reported in future publications.

{\it Acknowledgements}. This work was partially supported by
NSF-DMS-0204585, NSF-CAREER, and the Eppley Foundation for Research
(PGK); the Israel Science Foundation grant No.\ 8006/03 (BAM) and
the MECD/FEDER project BMF2003-03015/FISI (JC).

\newpage

\end{document}